\documentclass[useMAS,usenatbib]{mn2e}
\usepackage{natbib}
\usepackage{epsfig}
\usepackage{txfonts}
\usepackage{times}

\def\ms{\,m\,s$^{-1}$}         
\def\kms{\,km\,s$^{-1}$}         
\def\Msun{\hbox{$M_{\odot}$}}             
\def\Rsun{\hbox{$R_{\odot}$}}

\def\obj{CoRoT-22}
\def\corot{CoRoT}

\newcommand{\prob}[2]{p(\mathrm{#1}|\mathrm{#2})}

\begin{document}

\title[\obj\ b: a validated 4.9 R$_\oplus$ exoplanet in 10-day orbit]{\obj\ b: a validated 4.9 R$_\oplus$ exoplanet in 10-day orbit \thanks{The CoRoT space mission, launched on December 27th 2006, has been developed and is operated by CNES, with the contribution of Austria, Belgium, Brazil, ESA (RSSD and Science Program), Germany , Spain.}\thanks{Based on observations made with the HARPS instrument on the ESO 3.6 m telescope at La Silla (Chile), under the GTO program ID 072.C-0488 and the regular programs: 085.C-0019, 087.C-0831 and 089.C-0732.}}

\author[Moutou C. et al]{
Moutou, C.$^{1,2}$\thanks{E-mail: moutou@cfht.hawaii.edu}, 
 Almenara, J.M.$^2$,
 D\'{i}az, R.F.$^3$,
 Alonso, R.$^{4,25}$, 
 Deleuil, M.$^2$,
 Guenther, E.$^5$,\newauthor
 Pasternacki, T.$^6$, 
 Aigrain, S.$^7$,
 Baglin, A.$^8$,
 Barge, P.$^2$,
 Bonomo, A.S.$^9$,
 Bord\'e, P.$^{10}$,\newauthor
 Bouchy, F.$^2$,
 Cabrera, J.$^6$,
 Carpano, S.$^{11}$,
 Cochran, W.D.$^{12}$,
 Csizmadia, Sz.$^6$,
 Deeg, H.J.$^{4,25}$,\newauthor
 Dvorak, R.$^{13}$,
 Endl, M.$^{12}$,
 Erikson, A.$^6$,
 Ferraz-Mello, S.$^{14}$, 
 Fridlund, M.$^6$,
 Gandolfi, D.$^{15}$,\newauthor
 Guillot, T.$^{16}$,
 Hatzes, A.$^5$,
 H\'ebrard, G.$^{17}$,
 Lovis, C.$^3$,
 Lammer, H.$^{18,19}$,
 MacQueen, P.J$^{12}$,\newauthor
 Mazeh, T.$^{20}$,
 Ofir, A.$^{21}$,
 Ollivier, M.$^{10}$,
 P\"atzold, M.$^{22}$, 
 Rauer, H.$^{6,26}$,
 Rouan, D.$^8$,\newauthor
 Santerne, A.$^{23}$,
 Schneider, J.$^{24}$,
 Tingley, B.$^{4,25,27}$,
and Wuchterl, G.$^5$ \\
$^{1}$Canada France Hawaii Telescope Corporation, Kamuela,  USA]\\
$^{2}$Aix Marseille Universit\'e, CNRS, LAM (Laboratoire d'Astrophysique de Marseille) UMR 7326, Marseille, France\\
$^{3}$Obs. Astronomique de l'Univ. de Gen\`eve, Versoix, Switzerland \\
$^{4}$Instituto de Astrof\'{i}sica de Canarias, E-38205 La Laguna, Tenerife, Spain\\
$^{5}$Th\"uringer Landessternwarte, Tautenburg, Germany\\
$^{6}$Inst. of Planetary Research, German Aerospace Center, Berlin, Germany\\
$^{7}$Department of Physics, Oxford, UK\\
$^{8}$LESIA, Obs de Paris, Meudon , France\\
$^{9}$INAF - Osservatorio Astronomico di Torino, Pino Torinese, Italy\\
$^{10}$Institut d'Astrophysique Spatiale, Universit\'e Paris XI \& CNRS, 91405 Orsay, France \\
$^{11}$RSSD, ESTEC/ESA, Noordwijk, The Netherlands \\
$^{12}$McDonald Observatory, University of Texas, Austin, USA \\
$^{13}$University of Vienna, Institute of Astronomy, Vienna, Austria\\
$^{14}$IAG, Universidade de Sao Paulo, Brazil \\
$^{15}$Zentrum f\"ur Astronomie der Universit\"at Heidelberg, K\"onigstuhl 12, Heidelberg, Germany\\
$^{16}$Observatoire de la C\^ote d'Azur, Lab. Cassiop\'ee,  Nice, France\\
$^{17}$Observatoire de Haute Provence,  St Michel l'Obs, France\\
$^{18}$Institut d'Astrophysique de Paris, Paris, France\\
$^{19}$SRI, Austrian Academy of Science, Graz, Austria \\
$^{20}$School of Physics and Astronomy, R.B. Sackler Faculty of Exact Sciences, Tel Aviv, Israel \\ 
$^{21}$Institut f\"ur Astrophysik, Georg-August-Universit\"at, G\"ottingen, Germany \\
$^{22}$RIU an der Universit\"at zu K\"oln, Germany \\
$^{23}$Centro de Astrof\'{i}sica, Universidade do Porto, Rua das Estrelas, 4150-762 Porto, Portugal\\
$^{24}$LUTH, Obs de Paris, CNRS, Univ. Paris Diderot, Meudon, France\\
$^{25}$Universidad de La Laguna, Dept. Astrof\'{i}sica, E-38206 La Laguna, Tenerife, Spain\\
$^{26}$Zentrum f\"ur Astronomie und Astrophysik, Technische Universit\"at Berlin, 10623 Berlin, Germany\\
$^{27}$Stellar Astrophysics Center, Institut for Fysik og Astronomi, Aarhus Universitet, Ny Munkegade 120, 8000 Aarhus C, Denmark}
\date{Accepted 2014 August 10. Received 2014 August 5; in original form 2014 June 24}

\pagerange{\pageref{firstpage}--\pageref{lastpage}} \pubyear{2002}

\maketitle
\label{firstpage}

\clearpage
\begin{abstract}
The \corot\ satellite has provided high-precision photometric light curves for more than 163,000 stars and found several hundreds of transiting systems compatible with a planetary scenario. If ground-based velocimetric observations are the best way to identify the actual planets among many possible configurations of eclipsing binary systems, recent transit surveys have shown that it is not always within reach of the radial-velocity detection limits. 
In this paper, we present a transiting exoplanet candidate discovered by \corot\ whose nature cannot be established from ground-based observations, and where extensive analyses are used to validate the planet scenario. They are based on observing constraints from radial-velocity spectroscopy, adaptive optics imaging and the \corot\ transit shape, as well as from priors on stellar populations, planet and multiple stellar systems frequency. We use the fully Bayesian approach developed in the PASTIS analysis software, and conclude that the planet scenario is at least 1400 times more probable than any other false positive scenario. 
The primary star is a metallic solar-like dwarf, with M$_s$ = 1.099$\pm0.049$ \Msun\ and R$_s$ = 1.136$^{+0.038}_{-0.090}$ \Rsun . The  validated planet has a radius of R$_p$ = 4.88$^{+0.17}_{-0.39}$ R$_\oplus$ and mass less than 49 M$_\oplus$. Its mean density is smaller than 2.56 g.cm$^{-3}$ and orbital period is 9.7566$\pm$0.0012 days. This object, called \obj\ b, adds to a large number of validated Kepler planets. These planets do not have a proper measurement of the mass but allow statistical characterization of exoplanets population.
\end{abstract}

\section{Introduction}
\label{intro}
Photometry from space is currently  the most sensitive method to detect small-size planets with orbital periods less than a year. Twin super-earths CoRoT-7 b, Kepler-10 b and 55 Cnc e were the first objects in this category, and were detected each by a different space-based instrument \citep[CoRoT, $Kepler$, MOST/Spitzer, respectively, see][]{leger09,batalha10,winn11,demory11}. Although it was possible for the three examples cited above, measuring the mass of such small planets is in general difficult, requiring spectrographs of extreme radial-velocity precision. When the parent star is faint, as often for transiting systems, this mass characterization goes beyond the detection threshold. In such cases, one has to rely on alternative solutions to establish the transiting body as a planet. One is the detection of transit timing variations when several planets orbit the same star, as used for the Kepler-9 \citep{holman} and Kepler-11 \citep{lissauer} multiple systems. But this requires specific configurations that will not occur evenly in all systems. The next alternative is to examine and reject all other possible scenarios, with the aid eventually of additional observational constraints and intensive simulations. The $Kepler$ candidates \citep{borucki}  are in majority Neptune-like or super-earths, orbit stars generally fainter than V=14 with periods up to several months. There is only a small part of these for which high-precision radial-velocity measurements will allow the mass characterization. Undiluted binaries are easily rejected, but all configurations involving three bodies are more difficult to exclude. The $Kepler$ team is making use of the {\sc blender} analysis tool (e.g., \citet{torres2005,torres2011,fressin2011}) for some of individual candidates, by including a comparison of data with models. The outputs are the relative probability that a given system corresponds to a transiting planet or to a variety of other eclipsing system configurations. In the case of systems where hypothesis priors dominate, a statistical analysis not involving the data is also used with success \citep{lissauer14}. A mixture of both approaches (simplified data/model comparison and extended hypothesis priors from a model of the Galaxy) has also been provided for all $Kepler$ candidates by \citet{morton12}, giving an homogeneous look at the candidate validation process.
The observational constraints that are to be used in such cases are: the light curve and transit shape, the centroid time series (for $Kepler$ only), the high-resolution ground-based images of the star's vicinity, the radial-velocity (RV) times series, and the bisector-RV dependency.

In this paper, we present one such system discovered by CoRoT \citep{auvergne09, moutou13}. 
\obj\ b is a transiting candidate around a Sun-like star, for which a 9.75-day period transit has been detected by CoRoT in 2008. Intensive RV observations with HARPS and HIRES, adaptive-optics imaging with NACO and heavy false-positive simulations using PASTIS \citep{diaz14} have been performed in order to solve the system's nature. 

\begin{table}[h]
\caption{IDs, coordinates and magnitudes.}            
\centering        
\begin{minipage}[!]{7.0cm}  
\renewcommand{\footnoterule}{}     
\begin{tabular}{lcc}       
\hline\hline                 
CoRoT window ID & LRc02-E1-0591\\
CoRoT ID & 105819653\\
USNO-A2 ID  & $0900-13496507$ \\
2MASS ID   & $18424010+0613088 $ \\
\\
\multicolumn{2}{l}{Coordinates} \\
\hline            
RA (J2000)  & 18h42m40.12s \\
Dec (J2000) &  +06$^\circ$13'9.30''\\
\\
\multicolumn{3}{l}{Magnitudes} \\
\hline
\centering
Filter & Mag & Error \\
\hline
B$^a$  & 14.740 & 0.041	\\
V$^a$  & 13.944 & 0.083 \\
g'$^a$ & 14.299 & 0.041 \\
r'$^a$ & 13.698 & 0.084 \\	
i'$^a$ & 13.487 & 0.141 \\
J$^b$  & 12.414 & 0.027 \\
H$^b$  & 12.099 & 0.027 \\
Ks$^b$ & 11.988 & 0.026 \\
W1$^c$ & 11.851 & 0.026 \\	    
W2$^c$ & 11.917 & 0.026 \\
\hline\hline
\vspace{-0.75cm}
\footnotetext[1]{from APASS catalog (http://www.aavso.org/apass).}
\footnotetext[2]{from 2MASS catalog \citep{skrutskie}.}
\footnotetext[3]{from WISE catalog \citep{wright}.}
\end{tabular}
\end{minipage}
\label{startable}      
\end{table}

\section{Observations}\label{corot1}
\subsection{CoRoT}
\obj\ is the star referenced as 2MASS $18424010+0613088$ (Table~\ref{startable}). 
CoRoT observations\footnote{The data are publicly available on the \corot\ archive at http://idoc-corot.ias.u-psud.fr}  lasted from Julian date JD = 2454572.4656 (15 April 2008) to 2454717.4505 (7 September 2008), that is 145  days with a temporal bin of 32 seconds. The light curve consists of 336837 data points with a duty cycle of 87\%. 
Three light curves are extracted by the \corot\ pipeline, due to a low-dispersion prism in the light path. They correspond to the red, green and blue part of the stellar spectral energy distribution. The spatial aperture of each channel may also be differentially affected by contamination from neighbour stars.  An additional white light curve is extracted, summing all the flux in the photometric aperture.

A series of 0.2\% deep transits of periodicity 9.75 days is detected with $43\,\sigma$ significance in the white light curve.
To obtain clean colored light curves for modeling (Section~\ref{pastissec}), we removed the transits and then reproduced the low frequency variations of the residuals using a Savitzky-Golay smoothing filter. The resulting smoothed light curve was used to normalize the original light curve, removing the star's non-periodic activity signature and non-corrected instrumental effects. Finally, outliers were removed with a 3-$\sigma$ clipping performed on the folded transit.

\begin{figure}
\centering
\epsfig{file=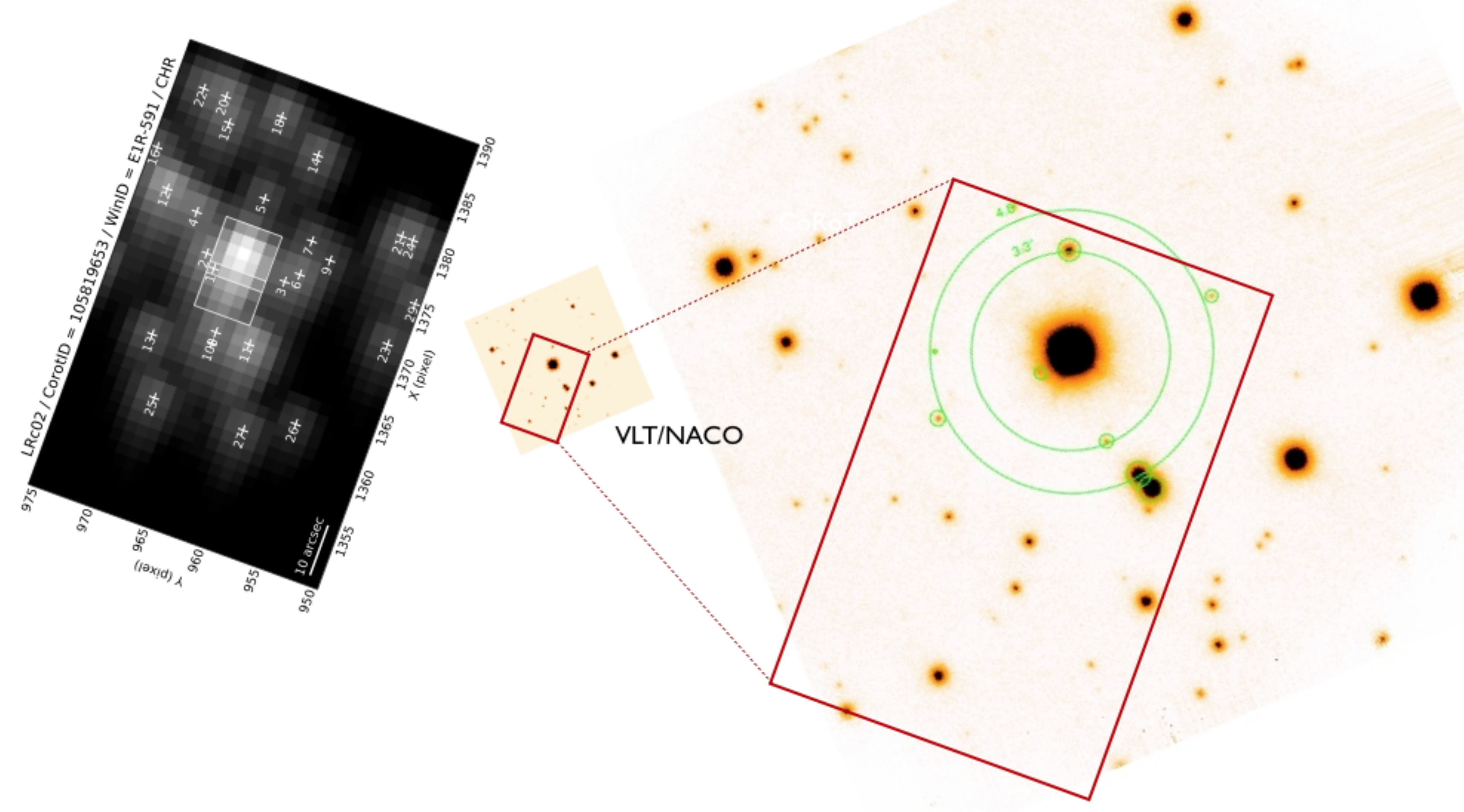,width=0.5\textwidth}
\caption{Left: Simulated image of the field surrounding the main target convolved by the PSF of \corot. Right: Field around the target \obj\ in the $J$ band as observed with VLT/NACO, with two zoom levels. The rectangle (white in the CoRoT image, red in the others) shows the aperture used for \corot\ photometry. The circles on the right image show the 3.3 and 4.8" radius distance from the main target, where the main nearby contaminants are located (see Figure 2). }
\label{fig1}
\end{figure}

Figure~\ref{fig1} shows the stellar neighborhood of \obj\ and the size and position of the \corot\ aperture mask. The flux arising from neighbor stars and the knowledge of the in-flight PSF allow to precisely estimate the contamination factor in all three chromatic channels \citep{borde2010}. The derived values for the rate of flux contamination are 1.9 $\pm$ 0.5 \% for the white channel. Contamination is larger in blue (6.2 $\pm$ 1.4 \%), moderate in green (4.1 $\pm$ 1.3 \%) and low in red  (1.0 $\pm$ 0.3 \%), with an expected impact on the transit depths.

\subsection{Ground-based photometry}
\label{phot3}
Imaging sequences from ground-based telescopes have been obtained for confirming the occurrence of the transit on the V=13.9 magnitude target \citep{deeg2009}. We used the IAC 80-cm telescope in Tenerife and the 1.2~m telescope Euler at La Silla to discard any source of photometric variation at the expected ephemeris in the neighborhood of the target. 

We conducted high-resolution, high-contrast observations using VLT/NACO\footnote{program ESO 285.C-5045} \citep{guenther13}. First, a J-band off-transit image was obtained, on 28 August 2010. Several more contaminating stars were detected in the high-quality off-transit images down to magnitude difference $\Delta J =$ 7 compared to the main target. Within the CoRoT aperture mask, one additional star is detected, at 3.3'' distance from the main target. It is extremely faint ($J \simeq 20$) and not detected in the optical images. Such star would have to get more than 200\% fainter to cause the eclipse, and is excluded as false positive.
Second, on-transit observations were performed with NACO on 7 October 2010. We performed aperture photometry with 0.4 arcsec radius aperture and did not detect changes larger than 5\% for all stars within the \corot\ mask. We measured on-off variations of all stars within the aperture of \corot\ as well as the variation they should have to induce the observed $\sim$0.2\% transit, assuming a constant color index from the $J$ band to the optical band of \corot\ observations. No variation compatible at $3\,\sigma$ with the expected one has been detected. This excludes all visible field star as the source of the transit, except the target itself.

\begin{figure} 
\centering
\vspace*{-1cm}
\includegraphics[width=0.5\textwidth]{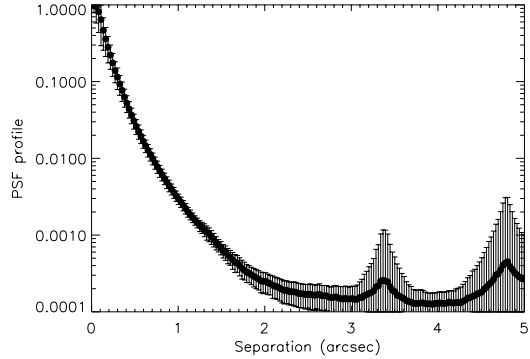}
\vspace*{-5cm}
\caption{NACO PSF profile around \obj\ estimated from  the NACO out-of-transit data. Bumps at 3.3 and 4.8" corresponds to contaminants depicted on Fig. 1.}
\label{fig3b}
\end{figure}

Figure~\ref{fig3b} shows the PSF profile close to the target, averaged over azimuthal rings of one pixel radius. It shows the detected stars as bumps, e.g. at 3.3". At shorter separation from the main target, we do not detect stars brighter than $2.5 \cdot 10^{-2}$, 3.3  $\cdot 10^{-3}$, 2.5 $\cdot 10^{-4}$ times than the target flux at respectively 0.5, 1.0 and 2.0". Such constraint (Figure 2) is used later for hypotheses prior calculations (section \ref{prior}).

From this multi-instrument search, we conclude that no detected stellar contaminant inside the \corot\ aperture of \obj\ produces a transit-like feature at the relevant time and depth, and thus resolved contaminating eclipsing binaries are safely ruled out. 

\subsection{Radial velocity}
\label{rv5}
\begin{figure}
\centering
\includegraphics[width=0.5\textwidth]{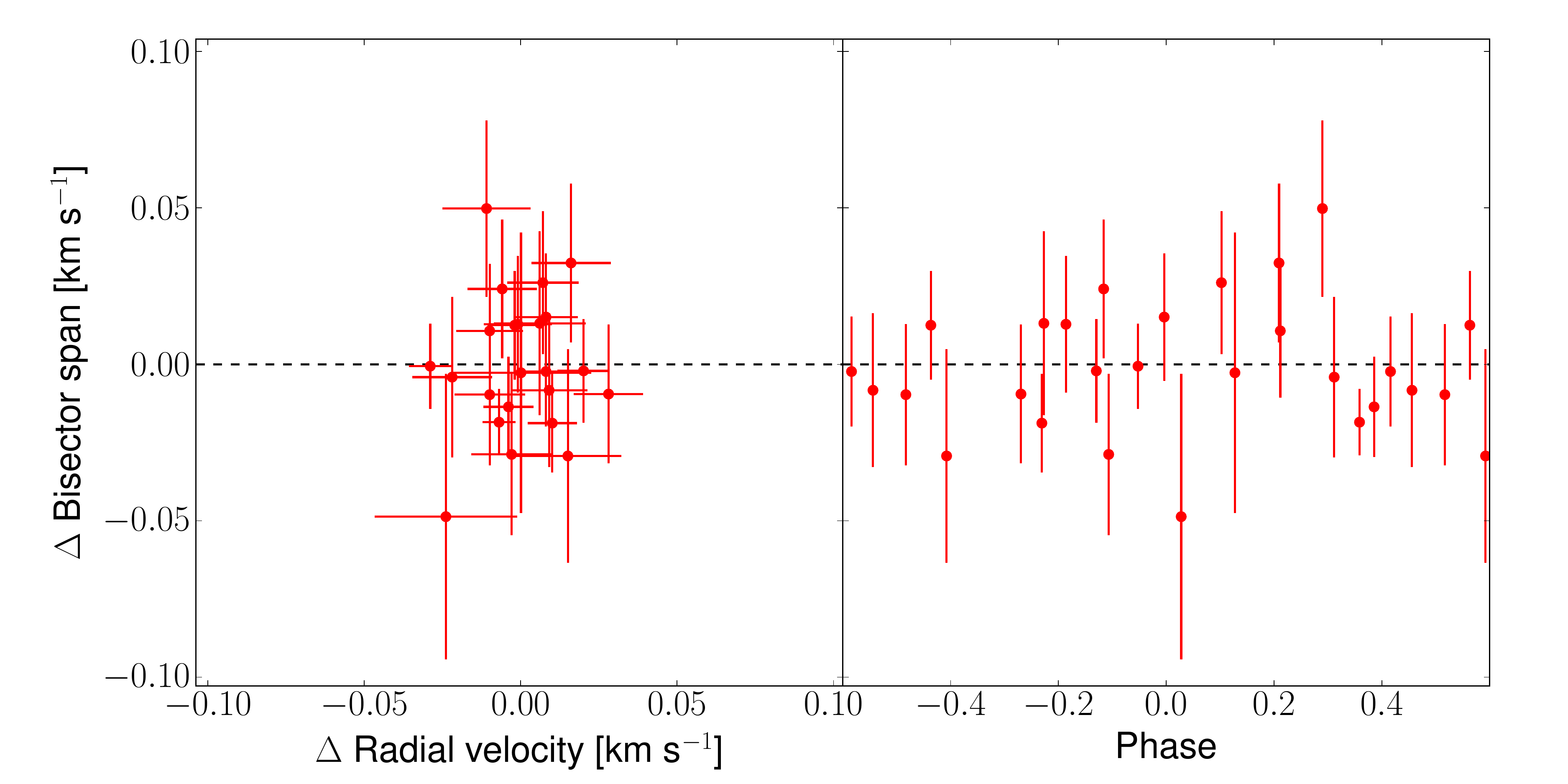}
\caption{HARPS bisector as a function of radial velocity (left panel) and as a function of orbital phase (right panel).}
\label{fig5}
\end{figure}

Three preliminary spectra were obtained with the OHP/SOPHIE spectrograph  in August 2008 \footnote{program PNP.MOUT}. They showed little variation in RV (less than 40 m/s with individual errors of 15 m/s), quickly excluding the undiluted binary scenario. We then secured 23 ESO-3.6/HARPS\footnote{ESO programs 072.C-0488, 083.C-0186 and 184.C-0639} \citep{mayor2003} measurements at La Silla, from September 2008 to August 2013. The RV are obtained by cross-correlating the spectra with a weighted numerical mask of G2 spectral type and by fitting the cross-correlation function with a Gaussian model.
Individual errors with HARPS are on average 10 m/s and the RV span a range of 42 m/s, with a standard deviation of 11 m/s. Average signal-to-noise ratio of the spectra is 15. The bisector slope ranges from -66 to +32 m/s over the HARPS sequence with typical errors of 20 m/s. Figure~\ref{fig5} shows the bisector span behavior as a function of RV, and as a function of the orbital phase. There is no correlation between the RV and the profile distortion. The non-detection of a second component in the cross-correlation function and the bisector behavior already put some constraints by excluding a bright blended star. 

In addition, 12 measurements were collected from June 2010 to July 2011 with the Keck1/HIRES spectrograph\footnote{Keck programs : N035Hr, N143Hr 260 and N095Hr} as part of the NASA key science project in support of the CoRoT mission. HIRES was used with the red cross-disperser, the 0.861" wide slit and the I$_2$-cell. The RV were derived with the $Austral$ Doppler code \citep{endl2000}. 
Table~\ref{rv} lists the RV obtained with the three spectrographs. 
The RV time series shows no significant variation at the expected \corot\ ephemeris (Fig. 4, top panel).

\begin{table}
\centering
\caption{Radial velocities, and bisector span.}
\begin{tabular}{lllll}
\hline
BJD-2400000. & RV & $\sigma_{RV}$ & Span & Instrument\\
days & \kms & \kms & \kms & \\
\hline
54705.36548 & 31.943 & 0.021 & -0.0276 & SOPHIE\\ 
54706.39550 & 31.918 & 0.015 & -0.0761 & SOPHIE\\
55006.47237 & 31.898 & 0.013 & -0.0351 & SOPHIE\\
\hline
55366.83337 & 0.9622 & 0.0069 &  & HIRES\\
55367.04615 & 0.9617 & 0.0071 &  & HIRES\\
55367.81985 & 0.9669 & 0.0069 &  & HIRES\\
55368.10482 & 0.9764 & 0.0065 &  & HIRES\\
55368.81888 & 0.9567 & 0.0084 &  & HIRES\\
55369.07215 & 0.9659 & 0.0054 &  & HIRES\\
55429.75227 & 0.9636 & 0.0099 &  & HIRES\\
55430.76439 & 0.956 & 0.011 &  & HIRES\\
55430.97373 & 0.9824 & 0.0082 &  & HIRES\\
55435.75115 & 0.9827 & 0.0050 &  & HIRES\\
55766.85109 & 0.9563 & 0.0041 &  & HIRES\\
55767.95179 & 0.9439 & 0.0043 &  & HIRES\\
\hline
54734.49357 & 31.8612 & 0.0068 & -0.0182 & HARPS\\ 
54742.54568 & 31.896 & 0.015 & -0.0045 & HARPS\\
54748.52255 & 31.8863 & 0.0080 & -0.0312 & HARPS\\
55020.75544 & 31.879 & 0.014 &  $\;$0.0322 & HARPS\\
55023.70810 & 31.905 & 0.017 & -0.0469 & HARPS\\
55323.86186 & 31.8829 & 0.0053 & -0.0361 & HARPS\\
55324.81150 & 31.900 & 0.012 & -0.0259 & HARPS\\
55325.85872 & 31.8878 & 0.0087 & -0.0051 & HARPS\\
55327.86799 & 31.8998 & 0.0079 & -0.0364 & HARPS\\
55328.85383 & 31.9097 & 0.0083 & -0.0197 & HARPS\\
55338.83366 & 31.887 & 0.013 & -0.0464 & HARPS\\
55339.83851 & 31.898 & 0.010 & -0.0025 & HARPS\\
55351.67286 & 31.906 & 0.013 &  $\;$0.0148 & HARPS\\
55353.69072 & 31.8977 & 0.0088 & -0.0199 & HARPS\\
55354.67423 & 31.880 & 0.011 & -0.0273 & HARPS\\
55359.65810 & 31.866 & 0.023 & -0.0663 & HARPS\\ 
55390.71984 & 31.880 & 0.011 & -0.0069 & HARPS\\
55391.69522 & 31.868 & 0.013 & -0.0217 & HARPS\\
55426.54670 & 31.885 & 0.011 & $\;$0.0065 & HARPS\\
55805.53299 & 31.918  &	0.011 & -0.0271 & HARPS\\
56472.81223 & 31.890  &	0.022 & -0.0203 & HARPS\\	
56511.59174 & 31.897  &	0.011 & $\;$0.0085 & HARPS\\	
56518.53339 & 31.889  &	0.011 & -0.0048 & HARPS\\
\hline
\end{tabular}
\label{rv}
\end{table}

\subsection{Spectroscopic parameters}
\label{spec4}
From multicolor photometric observations, we estimated that \obj\ is of  solar type and on the main sequence, with a best-fit template star of G3V type and low reddening ($E_{(B-V)}$ of 0.05). Spectroscopic observations (HARPS and HIRES, section \ref{rv5}) confirmed this result and were used to get accurate determination of stellar parameters. 
Using VWA \citep{bruntt}, we derive: an effective temperature of 5780~K $\pm$ 100~K, a surface gravity of 4.30 $\pm$ 0.10 and total metallicity 0.21 $\pm$ 0.10 dex, with a microturbulence velocity of 0.8 \kms. 

We estimate the stars' projected rotational velocity of 4.0 $\pm$ 1.5 \kms. The rotation period could be of the order of 16 days if the stellar rotation axis is perpendicular to the line of sight. There is no sign for chromospheric activity in the CaII H and K lines. The Lithium line at 607 nm is clearly detected with an abundance of 2.54. We excluded the pre-main sequence solutions with an age  20 -- 50 Myr, which is conflicting with the absence of activity and the low stellar rotation.

\section{PASTIS analysis}\label{pastissec}

Since the RV data presented above do not reveal a clear signal in phase with the transiting candidate, its planetary nature has to be confirmed by other means. Planet validation is a technique that compares the posterior probability of the planetary hypothesis against that of all possible false positive hypotheses. The ratio of the posterior probabilities between two competing hypotheses ($H_i$ and $H_j$) is called the odds ratio:
\begin{equation}
O_{ij} = \frac{\prob{H_i}{D, I}}{\prob{H_j}{D, I}} = \frac{\prob{H_i}{I}}{\prob{H_j}{I}}\cdot\frac{\prob{D}{H_i, I}}{\prob{D}{H_j, I}}\;\;,
\label{eq.oddsratio}
\end{equation}
where $D$ represents the data, and $I$ is the prior information available. The first term on the right-hand side of the equation is called the hypotheses priors ratio, or prior odds, and depends exclusively on $I$. The second term is called the Bayes factor. All the support the data provide to a given hypothesis is contained in the Bayesian evidence $\prob{D}{H_i, I}$. The Bayes factor is simply the ratio of evidences for the two competing hypotheses.
The computation of the likelihood is done with the following assumptions: i) the data follow a normal distribution; ii) the datasets are independent.

We computed the odds ratio between the planet hypothesis and four false positive hypotheses using the Planet Analysis and Small Transit Investigation Software \citep[PASTIS,][]{diaz14}. We refer the reader to this paper for more details on the method, and performance analyses. The scenarios we considered are: i) diluted eclipsing binary (BEB), where the eclipsing binary is aligned by chance with the target, ii) diluted transiting planet (BTP), where a transiting planet orbits a star aligned by chance with the target, iii) planet in binary (PiB), where the target is a (non-eclipsing) binary and the planet orbits one component of the system, and iv) hierarchical triple system (TRIPLE), where the main target is composed by three stars, two of which being eclipsing. Scenarios where the observed transit is the secondary eclipse of a diluted binary are also automatically considered  by PASTIS. The false positives hypotheses involving periods twice the nominal transit period poorly fit the data and are no longer considered.\\

{\bf Data:} The observational data used are all RV measurements (Table~\ref{rv}), CoRoT red, green, blue, and total light curves (Section~\ref{corot1}), and the magnitudes in filters as listed in Table~\ref{startable}. The RV and the light curves are fitted in time to propagate the error on the ephemerides to the rest of parameters. 

{\bf Stellar models:} As stellar models we used the Dartmouth stellar tracks \citep{dotter}, the PHOENIX/BT-Settl stellar atmosphere models \citep{allard}, together with the limb-darkening coefficients from the tables by \citet{claret} computed for CoRoT. For the CoRoT red, green, and blue colors, the limb-darkening coefficients are  interpolated from the limb-darkening coefficients for Sloan filters taking into account the overlap with a given CoRoT band and the star emission. The light curves were computed using the EBOP code \citep{nelson, etzel, popper} extracted from the JKTEBOP package \citep{southworth}. As the wavelength range covered by each CoRoT band changes from one target to another, we multiply the SED modeled in each MCMC step by the CoRoT full instrument response. Then we divided this spectrum in three bands (red, green and blue) to match the relative flux contribution in the light curve, after correcting the flux for the contamination (as given in Section~\ref{corot1}). \\

The following free parameters are common to all hypotheses: i) transit ephemeris; ii) spectrographs offsets; iii) spectrographs, light curves, and SED systematic white noise amplitude; iv) light curve contamination (Section~\ref{corot1}); v) out-of-transit light-curve flux normalization factor. Table~\ref{prior_table} gives the prior distributions for all parameters of all models.

{\bf Stars:} In all hypotheses except PLANET, the main star in the CoRoT photometric mask, also called the target star, is modeled with its effective temperature ($T_{\mathrm{eff}}$), surface gravity (log\,$g$), and metallicity ($[\rm{Fe/H}]$) with priors from the values obtained in the spectroscopic analysis (Section~\ref{spec4}, Table~\ref{prior_table}). In the PLANET hypothesis, the parameters describing the host star are $T_{\mathrm{eff}}$, log\,$g$, and stellar density ($\rho_{\star}$), as the latter is constrained by the transit \citep[see][]{diaz13}; its prior is estimated from $T_{\mathrm{eff}}$, log\,$g$, $[\rm{Fe/H}]$, and the Dartmouth stellar tracks.
Other stars in the hypotheses are modeled by their age, $[\rm{Fe/H}]$, and initial mass. If two or three stars are physically bounded they share the same age and $[\rm{Fe/H}]$. The prior for the age and the $[\rm{Fe/H}]$ are uniform distributions covering the full range of the models, except for the upper limit of the age that is limited by the age of the Universe. The prior for the mass comes from the initial mass function for stars of the disk of the Galaxy from \citet{robin2003}. The albedo of each star is a free parameter with uniform prior between 0.6 and 1.0 \citep{claret01}; see Table~\ref{prior_table} for a detailed version of each prior. 

{\bf Planets:} In the PLANET hypothesis, the planet is modeled by the radius ratio, and the a/R$_s$ is computed from $\rho_{\star}$, the period, and the mass ratio. In all other hypotheses, the planets are modeled by their mass and radius with uniform priors (Table~\ref{prior_table}). All planets have the albedo as a free parameter with uniform priors between 0 and 1. We are therefore also exploring the possibility of diluted transiting brown dwarfs, although they are considered as non-emitting bodies.

{\bf Orbits and RV:} We used uniform priors for the eccentricity $e$, argument of periastron $\omega$, and impact parameter $b$. The amplitude of the RV in the PLANET scenario is a free parameter with uniform priors, while for other scenarios it is computed from the components' masses and the orbital parameters. A linear drift is fitted in all cases, with uniform prior. The second, wider orbit of the PiB and TRIPLE scenarios is considered to have a period long enough so that the effects in the RV are negligible. Then the velocity of the respective component or couple is a free parameter. 

{\bf Distance:} The distance to the target star and the diluted systems is a free parameter with a prior uniform in d$^2$, obtained by considering that the stars are uniformly distributed in the sky. The diluted systems are forced to be at least one magnitude fainter than the target star; brighter stars are easily excluded from the current data set. The extinction is modeled using the three-dimensional extinction model of \citet{amores}, that depends on Galactic coordinates and distance. We used the  extinction law from \citet{fitzpatrick} with $R_{\mathrm{V}}=3.1$. \\

We have run 50 MCMC chains of $3\times10^5$ steps for each hypothesis. The chains were started at random points drawn from the joint prior distribution, except for the main star in the CoRoT photometric mask whose parameters are started at the most probable value to reduce the burn-in interval.

\begin{figure*}
\centering
\includegraphics[height=3.5cm]{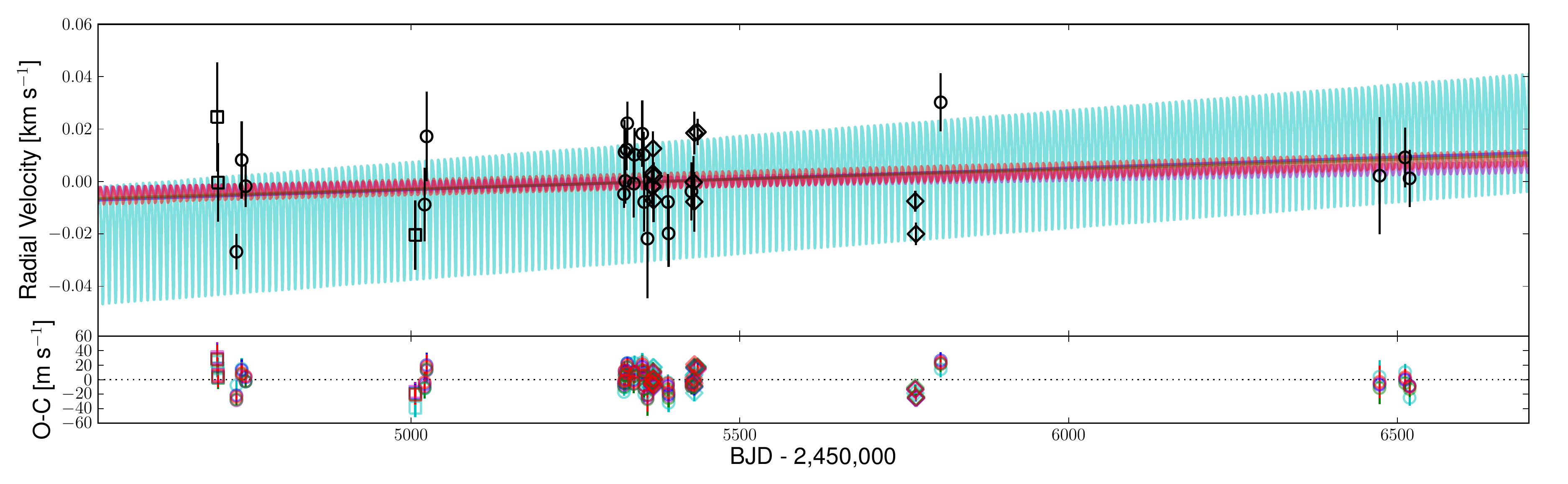}\includegraphics[height=3.5cm]{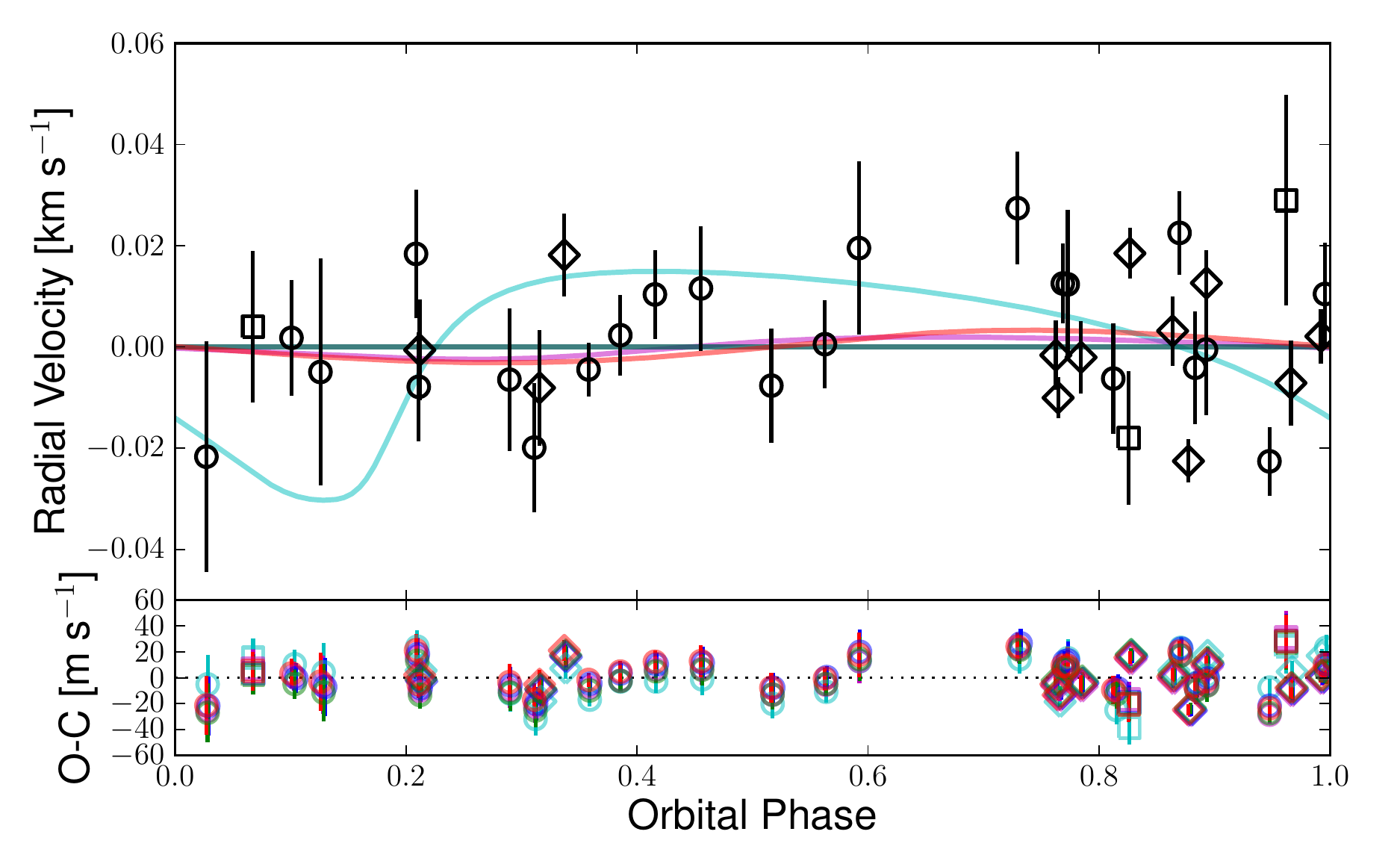}\\
\includegraphics[height=3.5cm]{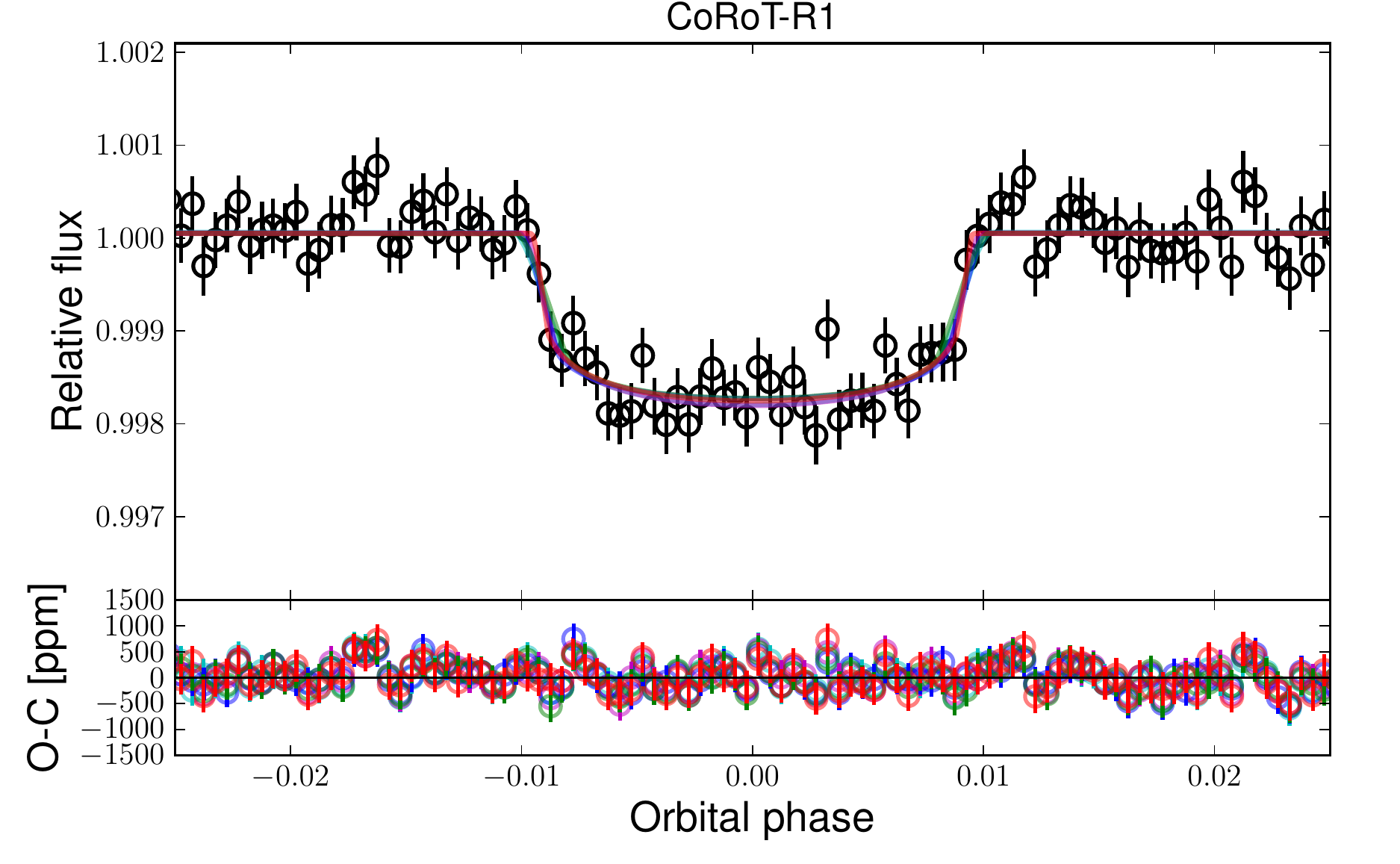}\includegraphics[height=3.5cm]{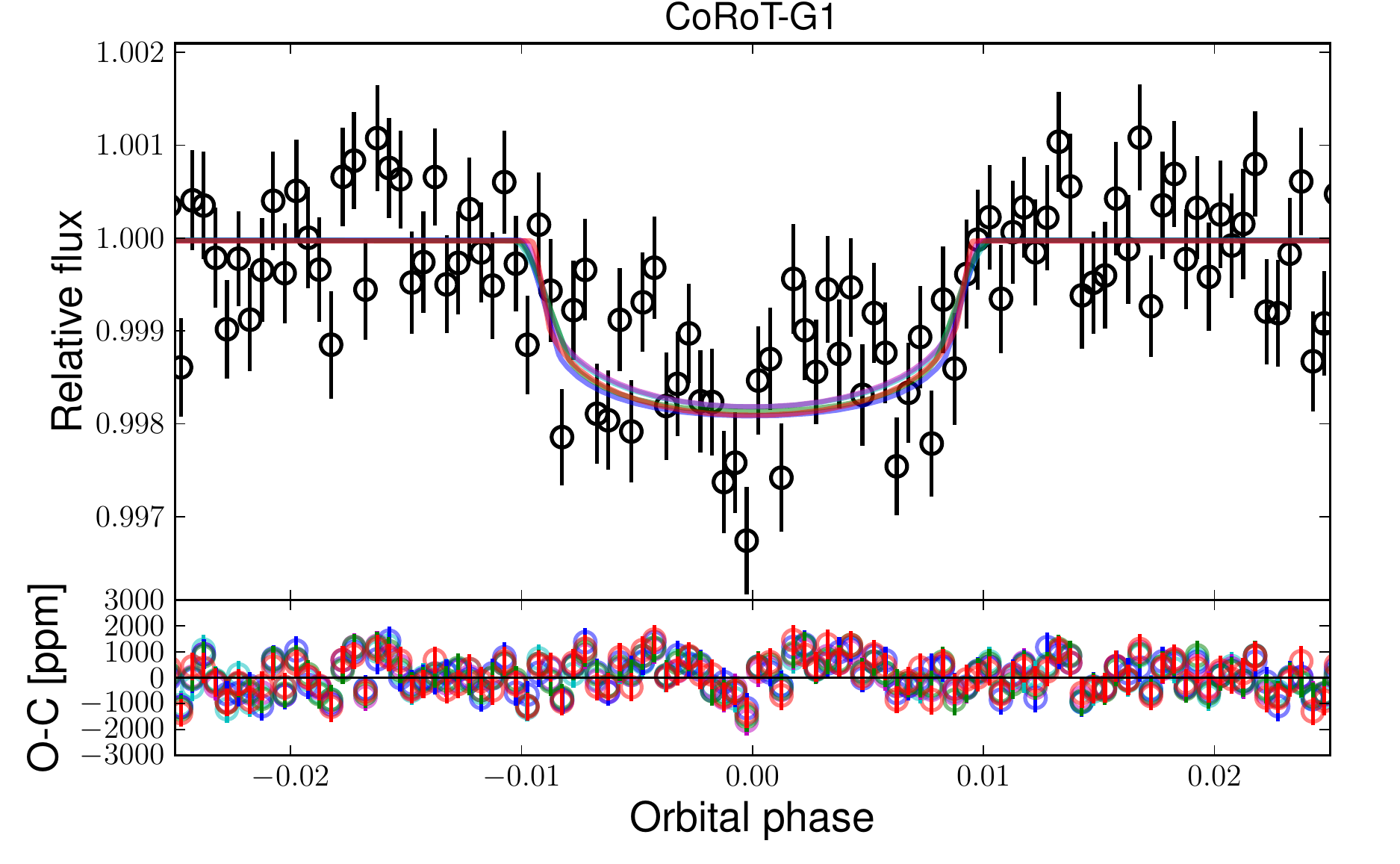}\includegraphics[height=3.5cm]{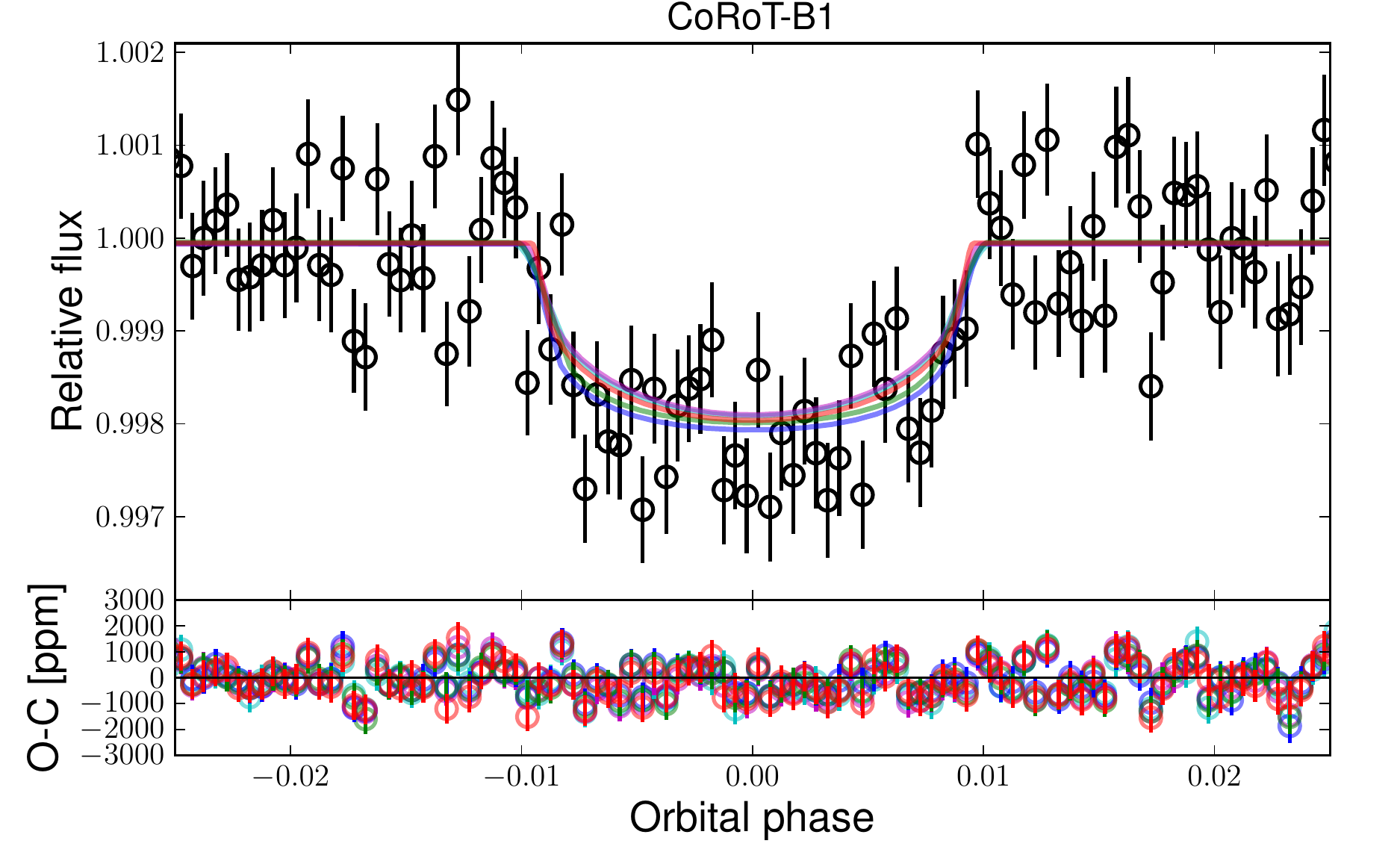}\\
\includegraphics[height=3.5cm]{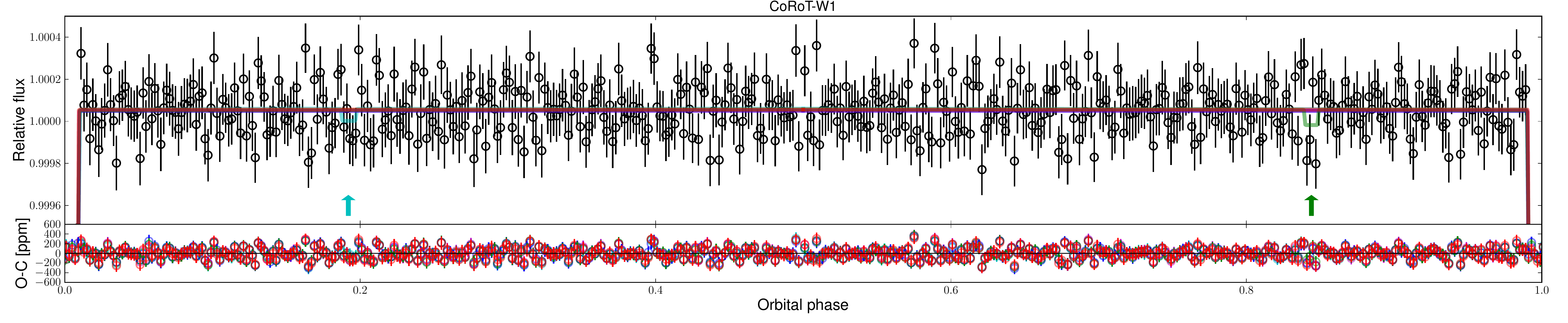} \\
\includegraphics[height=3.5cm]{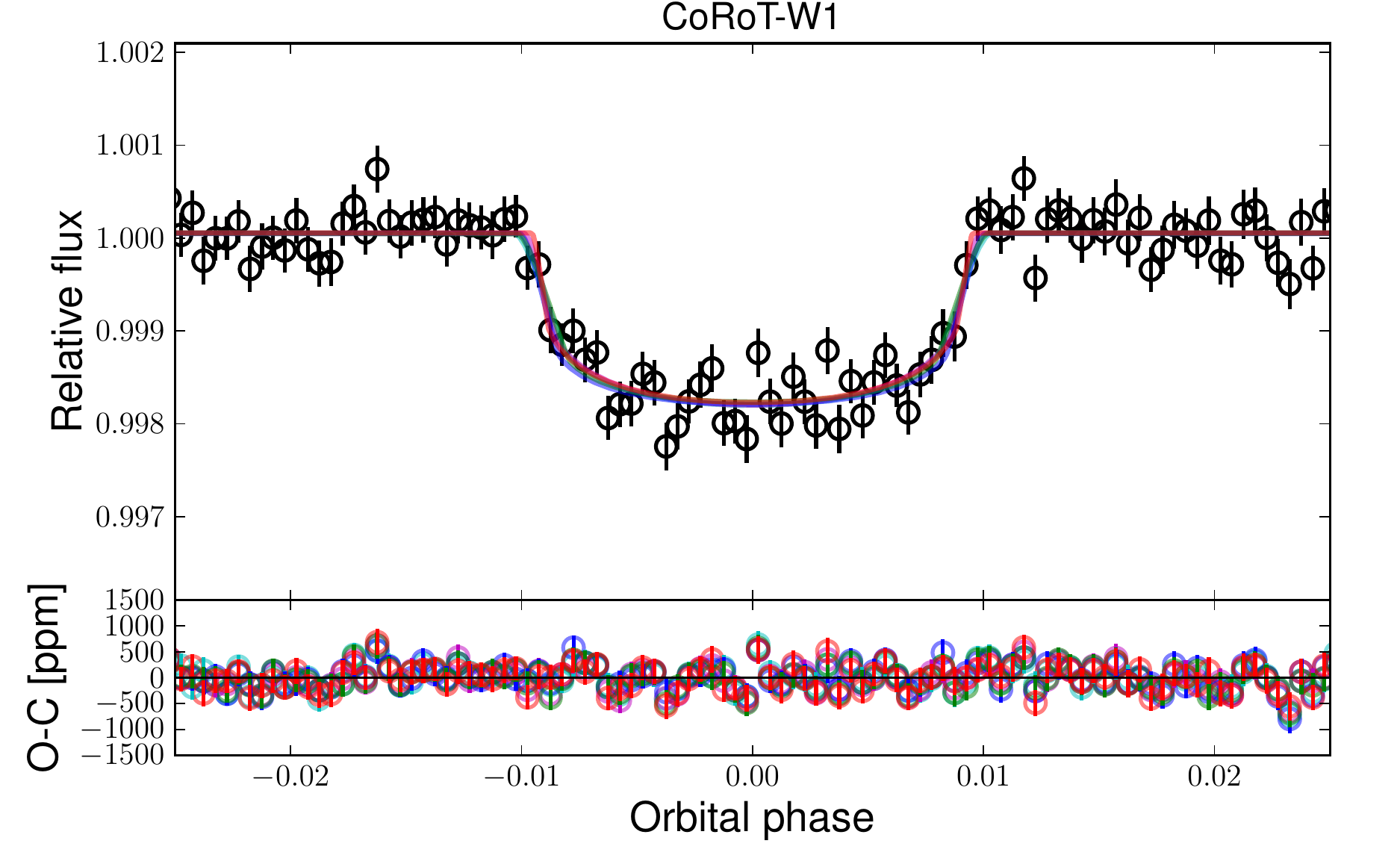}\includegraphics[height=3.5cm]{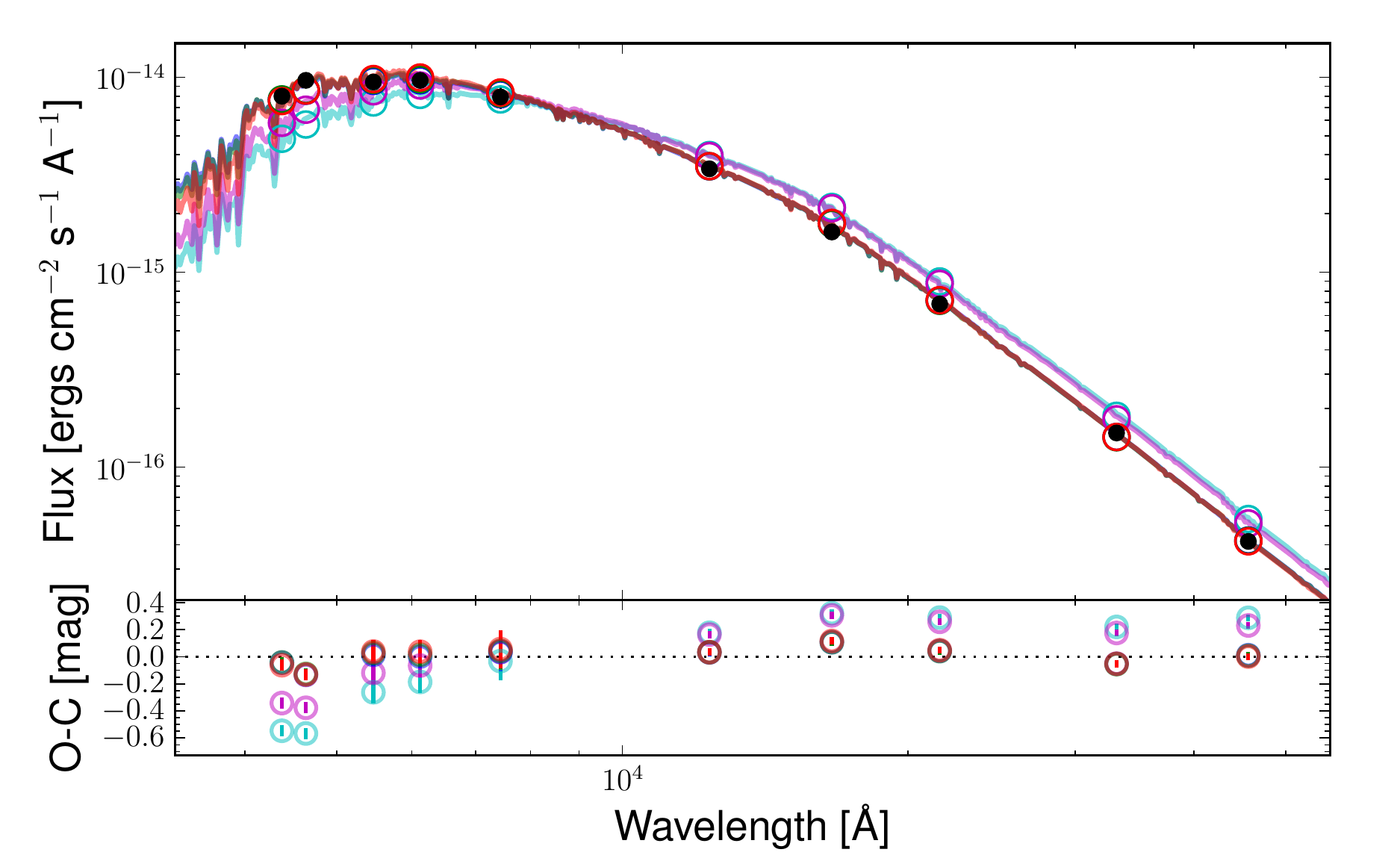}\includegraphics[height=3.5cm]{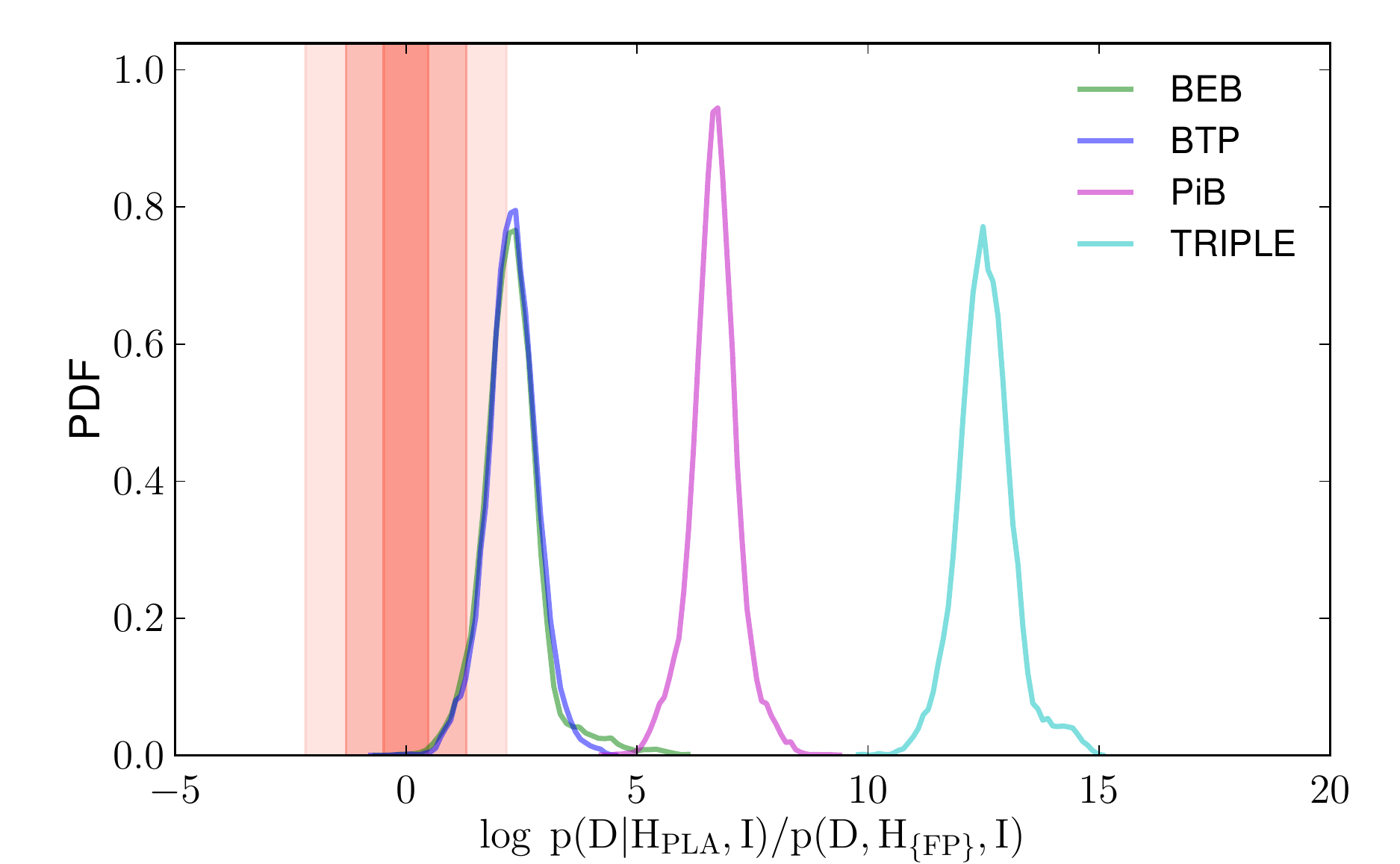}\\
\caption{From top to bottom, and left to right: i) Radial-velocity data set against time, HARPS as open circles, HIRES as open diamonds, SOPHIE as open squares; superimposed is the best-fit model for the five different hypotheses: PLANET (red), BEB (green), BTP (blue), PiB (magenta), and TRIPLE (cyan). The residuals to the models are shown in the bottom plot. ii) Radial-velocity data set against orbital phase (any linear drift subtracted). iii) CoRoT red, green, and blue light curves folded in phase and binned. iv) The out-of-transit light-curve showing the secondaries of the BEB and TRIPLE hypotheses with an arrow. v) The white light curve, plotted for all phases and zoomed around the transit. vi) Spectral energy distributions (solid lines), with the flux integrated in each of the photometric bands(open circles), and observed magnitudes (black dots). vii) Probability distribution function for the Bayes factor between the planet hypothesis and each false positive hypothesis (see text). }
\label{figblend}
\end{figure*}

Figure~\ref{figblend} shows the data with the maximum-posterior model found for each of the hypotheses. Table~\ref{starplanet_param_table} shows the mode and the 68.3\% central confidence interval of the MCMC distributions for the PLANET hypothesis. 
  
\begin{table*}
\vspace{0cm}
\centering
\caption{Results of the PLANET hypothesis fit.}            
\vspace{0cm}
\begin{minipage}[t]{13.0cm} 
\setlength{\tabcolsep}{10.0mm}
\renewcommand{\footnoterule}{}                          
\begin{tabular}{l l}        
\hline\hline                 
\vspace{-0.25cm}\\
Planet orbital period, $P$ [days]$^{\bullet}$       & 9.75598 $\pm$ 0.00011 \\
Midtransit time, $T_{c}$ [BJD]$^{\bullet}$          & 2454598.42014$^{+0.00099}_{-0.00051}$ \\
$cov(P,T_{c})$ [days$^2$]                           & -4.28e-8     \\
Orbital eccentricity, $e$$^{\bullet}$               & 0.077$^{+0.30}_{-0.042}$, $\textless$ 0.25, $\textless$ 0.78$^{\dagger}$ \\
Argument of periastron, $\omega$ [deg]$^{\bullet}$  & 210$^{+70}_{-30}$   \\
Orbit inclination, $i$ [deg]                       & 89.749$^{+0.032}_{-0.90}$  \\
Orbital semi-major axis, $a$ [AU]                  & 0.0920 $\pm$ 0.0014   \\
semi-major axis / radius of the star, $a/R_{\star}$ & 17.30$^{+1.3}_{-0.57}$ \smallskip \\
Radius ratio, $k=R_{p}/R_{\star}$$^{\bullet}$         & 0.03927$^{+0.00060}_{-0.00037}$ \\
Transit duration [h]                      & 4.492 $\pm$ 0.039 \\
Impact parameter, $b$$^{\bullet}$                    & 0.012$^{+0.33}_{-0.033}$\smallskip \\

Radial velocity linear drift [\ms year$^{-1}$]$^{\bullet}$ & 2.2$^{+2.1}_{-3.2}$ \\
Radial velocity semi-amplitude, $K$ [\ms]$^{\bullet}$ & 3.7 $\pm$ 3.9, $\textless$ 5.6, $\textless$ 15.2$^{\dagger}$ \smallskip\\

Contamination CoRoT-W, $C_{CoRoT-W}$ [\%] $^{\bullet}$ & 1.72 $\pm$ 0.57 \\
Contamination CoRoT-R, $C_{CoRoT-R}$ [\%] $^{\bullet}$ & 1.09$^{+0.18}_{-0.43}$ \\
Contamination CoRoT-G, $C_{CoRoT-G}$ [\%] $^{\bullet}$ & 4.4$^{+1.5}_{-1.0}$ \\
Contamination CoRoT-B, $C_{CoRoT-B}$ [\%] $^{\bullet}$ & 5.4 $\pm$ 1.4\smallskip\\
Offset SOPHIE, $\Delta RV_{SOPHIE}$ [\kms]$^{\bullet,\S}$ & -31.921$^{+0.017}_{-0.027}$ \\
Offset HIRES, $\Delta RV_{HIRES}$ [\kms]$^{\bullet,\S}$   & -0.9638 $\pm$ 0.0047 \\
Offset HARPS, $\Delta RV_{HARPS}$ [\kms]$^{\bullet,\S}$   & -31.8910$^{+0.0039}_{-0.0024}$\smallskip\\
Jitter CoRoT-W, $\sigma_{CoRoT-W}$ [mmag] $^{\bullet}$ & 0.9238$^{+0.0097}_{-0.020}$\\
Jitter CoRoT-R, $\sigma_{CoRoT-R}$ [mmag] $^{\bullet}$ & 1.064 $\pm$ 0.022 \\
Jitter CoRoT-G, $\sigma_{CoRoT-G}$ [mmag] $^{\bullet}$ & 2.237$^{+0.022}_{-0.038}$ \\
Jitter CoRoT-B, $\sigma_{CoRoT-B}$ [mmag] $^{\bullet}$ & 2.348 $\pm$ 0.034 \\
Jitter SOPHIE, $\sigma_{SOPHIE}$ [\ms]$^{\bullet}$     & 15.0$^{+46}_{-5.2}$ \\
Jitter HIRES, $\sigma_{HIRES}$ [\ms]$^{\bullet}$       & 10.08$^{+6.6}_{-0.90}$ \\
Jitter HARPS, $\sigma_{HARPS}$ [\ms]$^{\bullet}$       & 8.9$^{+4.2}_{-2.7}$ \\
Jitter SED, $\sigma_{SED}$ [mag]$^{\bullet}$           & 0.065$^{+0.056}_{-0.018}$ \smallskip\\

Star effective temperature, $T_{\mathrm{eff}}$[K]$^{\bullet}$   & 5939$^{+54}_{-120}$ \\
Star metallicity, $[\rm{Fe/H}]$ [dex]$^{\bullet}$             & 0.170$^{+0.070}_{0.12}$ \\ 
Star density, $\rho_{\star}$ [$\rho_\odot$]$^{\bullet}$ & 0.729$^{+0.19}_{-0.071}$ \\
Star mass, $M_\star$ [\Msun]                             & 1.099 $\pm$ 0.049 \\
Star radius, $R_\star$ [\Rsun]                           & 1.136$^{+0.038}_{-0.090}$ \\
Deduced stellar surface gravity, $\log$\,$g$ [cgs]      & 4.361$^{+0.065}_{-0.029}$ \\
Age of the star, $t$ [$Gyr$]                            & 3.3 $\pm$ 2.0 \\
Distance of the system, d [pc]$^{\bullet}$                 & 592$^{+25}_{-42}$ \\
Planet mass, $M_{p}$ [M$_\oplus$ ]                       & 12.2$^{+14}_{-8.8}$, $\textless$ 19.0, $\textless$ 48.9$^{\dagger}$ \\
Planet radius, $R_{p}$[R$_\oplus$]                       & 4.88$^{+0.17}_{-0.39}$ \\
Planet mean density, $\rho_{p}$ [$g\;cm^{-3}$]          & 0.249$^{+1.0}_{-0.097}$, $\textless$ 0.94, $\textless$ 2.56$^{\dagger}$ \\ 
Planet surface gravity, $\log$\,$g_{p}$ [cgs]          & 2.88$^{+0.12}_{-0.64}$, $\textless$ 2.89, $\textless$ 3.30$^{\dagger}$\\
Planet equilibrium temperature$^\ast$, $T_{eq}$ [K]     & 885$^{+53}_{-280}$ \smallskip\\
\hline       
\vspace{-0.5cm}
\end{tabular}
\begin{list}{}{}
\item $^{\bullet}$ MCMC jump parameter
\item $^{\dagger}$ upper limit at 68.3\%, and 99\% confidence
\item $^{\S}$ reference time BJD 2455379
\item $^{\ast}$ $T_{eq}=T_{\mathrm{eff}}\left(1-A\right)^{1/4}\sqrt{\frac{R_\star}{2 a}}$, using an albedo $A=0$.
\end{list}
\end{minipage}
\label{starplanet_param_table}  
\end{table*}

\subsection{Bayes factor}
To compute the Bayes factor, PASTIS estimates the evidence with the truncated posterior-mixture estimation \citep[TPM,][]{tuomi}. 
TPM is known to exhibit some issues concerning the estimation of the Bayes factor. Mainly, it does not penalise models sufficiently for each extra parameter. However, this gives a conservative odds ratio in favour of the planet, which is the model with fewer parameters.

The distribution of the Bayes factors are shown on Figure~\ref{figblend} (bottom right). On this plot, the shaded areas indicate the regions where the support of one hypothesis over the other is considered ''Inconclusive'', ''Positive'', ''Strong'', and ''Very strong'' (from darker to lighter region, and finally white), according to the criteria of \citet{Kass}.
The mode and the 68.3\% central confidence interval of the probability distribution function of the Bayes factor are given in Table~\ref{hypodds} for each scenario compared to the PLANET scenario. 


\begin{table*}
\begin{center}
\caption{Bayes factor, hypotheses priors, hypotheses priors ratios, odds ratios, and posterior.}
\begin{tabular}{lccccc}
\hline
\vspace{-0.25cm}\\
Hypothesis ($j$) & $\log\frac{\prob{D}{H_{PLANET}, I}}{\prob{D}{H_j, I}}$ & $\prob{H_j}{I}$ & $\frac{\prob{H_{PLANET}}{I}}{\prob{H_j}{I}}$ & $\log\frac{\prob{H_{PLANET}}{D, I}}{\prob{H_j}{D, I}}$ & $\log\prob{H_j}{D, I}$ \smallskip\\
\hline
\vspace{-0.25cm}\\
PLANET & 0				 & $1.2\times10^{-8}$ 	& 1 & 0				 & $-0.000138_{2.5e-4}^{8.3e-5}$ \\
BEB    & $2.30^{+0.37}_{-0.71}$ & $3.2\times10^{-10}$ & 37 & $3.87^{+0.37}_{-0.71}$ & $-3.77 \pm 0.54$ \smallskip\\
BTP    & $2.32^{+0.40}_{-0.64}$ & $1.6\times10^{-10}$ & 72 & $4.18^{+0.40}_{-0.64}$ & $-4.10 \pm 0.53$ \smallskip\\
PiB    & $6.70^{+0.33}_{-0.58}$ & $2.6\times10^{-9}$  & 4.5 & $7.35^{+0.33}_{-0.58}$ & $-7.31 \pm 0.60$ \smallskip\\
TRIPLE & $12.44 \pm 0.56$ & $2.0\times10^{-9}$  & 5.8 & $13.20 \pm 0.56$ & $-13.24 \pm 0.56$ \smallskip\\
\hline
\end{tabular}
\label{hypodds}
\end{center}
\end{table*}

The PLANET hypothesis is 200, 209, $5.0\times10^6$, and $2.8\times10^{12}$ times more probable than the BEB, BTP, PiB and TRIPLE hypotheses respectively, based only on the data and parameter priors. The TRIPLE and PiB hypotheses are rejected by the SED data, because such systems would have different relative magnitudes, especially in the blue; the diluted scenarios, however, are not excluded by the data, at this point.

\subsection{Hypotheses priors ratios}
\label{prior}
	The hypotheses prior ratios are computed as described in \citet[][Sect.~5]{diaz14}. Basically, we simulated the stellar field around the target using the Besan\c con Galactic model \citep{robin2003}, and computed the probability that an unseen blended star lies within a certain distance of the target, given the NACO contrast curve (section \ref{phot3}). The binary properties from \citet[][and references therein]{raghavan2010} are used to compute the probability that any such contaminating star is an eclipsing binary system; triple system statistics comes from \citet{rappaport13}. The giant planet statistics from \citet{fressin2013} and \citet{bonfils2013} are used to compute the probability that a star in the background hosts a transiting planet. Combining these probabilities, we computed the priors for the BEB and BTP hypotheses. The same sources are used to compute the prior probabilities of the TRIPLE hypothesis, the PLANET hypothesis, and the PiB hypothesis. For the latter, we further assumed that the presence of a planetary companion in orbit around one of the components of a wide-orbit binary is independent of the presence of the stellar companion and we arbitrarily imposed that the period of the stellar pair is at least 10 times the period of the planet around the secondary. If ones includes NACO observing constraints into the prior calculation for the PiB scenario, one gets a PiB probability value about twice smaller than when one does not, and both values are compatible within errors. This scenario having a seven orders of magnitude lower probability that the PLANET scenario, we conclude that, in the case analyzed here, taking into account the NACO constraint into the non-diluted PiB scenario is not necessary. Same reasoning applies for the TRIPLE scenario.
The hypotheses priors and ratios are listed in Table~\ref{hypodds}.



\subsection{Odds ratios and planet posterior probability}
\begin{figure*}
\centering
\includegraphics[width=5cm]{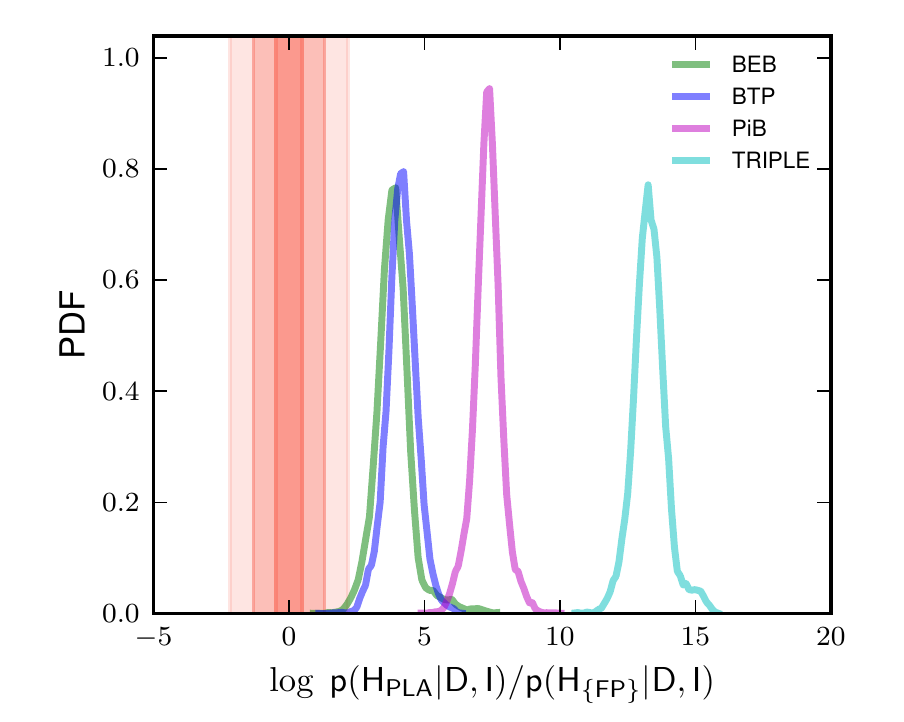}
\includegraphics[width=5cm]{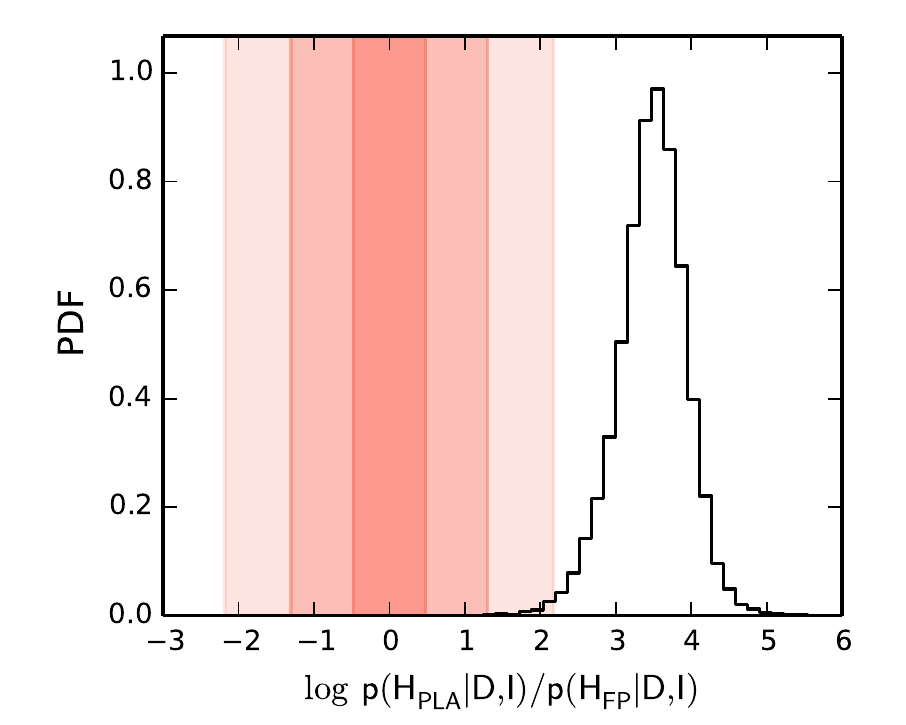}
\includegraphics[width=5cm]{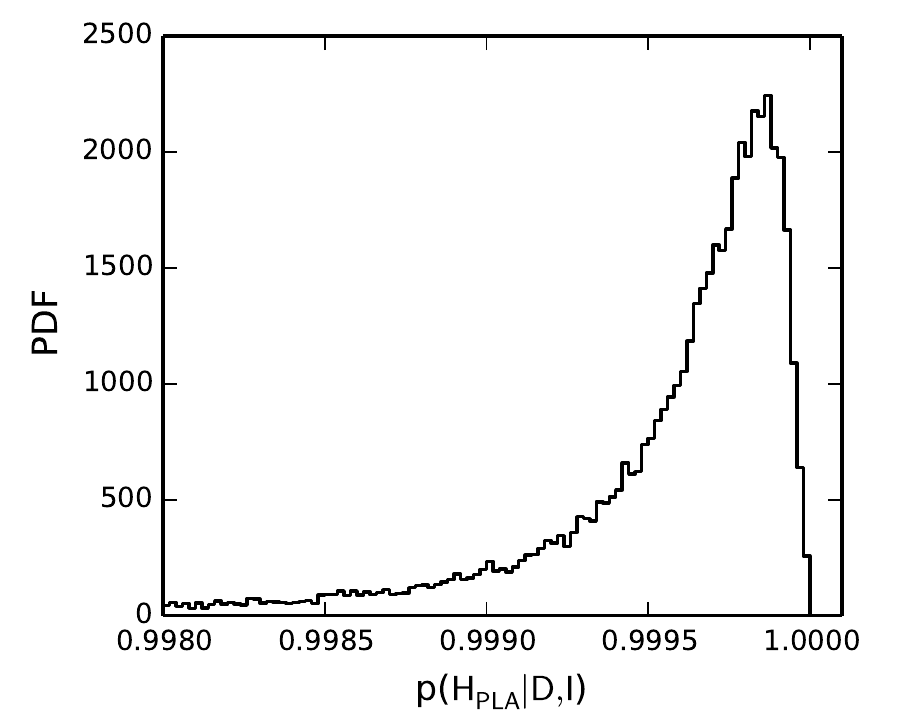}
\caption{From left to right: Odds ratios between the PLANET and the false positives hypotheses. Odds ratio between the PLANET hypothesis and any false positive hypothesis. Posterior probability of the PLANET hypothesis.}
\label{oddsratio_posterior}
\end{figure*}

The odds ratios of the planet hypothesis and each false positive hypothesis is computed as the product of the Bayes factor and the hypothesis prior ratio. The results are plotted in Figure~\ref{oddsratio_posterior} (left panel) and listed in Table~\ref{hypodds} for each hypothesis. The odds ratio between the PLANET hypothesis and any false positive hypothesis is $\log_{10}(\prob{\mathrm{PLANET}}{I; D}/\prob{\mathrm{FP}}{I; D}) = 3.50 \pm 0.46$, where FP is the hypothesis that the candidate is any false positive, independently of its kind. Figure~\ref{oddsratio_posterior} (middle panel) shows the distribution of the odds ratio. Around 99.24\% (resp. 96.58 \%) of the mass of the distribution is above 150 (resp. 370. i.e. the inverse probability of being outside $3-\sigma$ in a normal distribution).


Under the assumption that the set of hypotheses tested is complete, and that the sum of the posterior probabilities is, therefore, one, the odds ratio can be recast to obtain the posterior probability of the planetary hypothesis. The resulting distribution is shown in Figure~\ref{oddsratio_posterior} (right panel). The planet hypotheses posterior probability is larger than 0.994 (resp. 0.976) with 99.0\% (resp. 99.9\%) confidence. The mode of the distribution is 0.999856. 

Hypothesis priors alone showed that the planet scenario is the most likely solution, but with only a small advantage, as the planet/FP ratio ranges from 4 to 72. The scenarios favored by priors are, however, the one most easily discarded by the data (PiB and TRIPLE). Final odds ratios range between 1400 and 10$^{13}$ in favor of the planet scenario. The most likely scenario after the planetary system is a BEB, at least 1400 times less probable.

We conclude that, based on current data and prior knowledge, \obj\ b is validated as a planet with a very high degree of confidence.

\section{Conclusion}
\label{discu8}
In this paper, we report the discovery of a transiting planetary system, \obj. The mass of the planet could not be measured through Doppler measurements. With the decreasing size of transiting candidates, and the faintness of their parent stars, as provided by the photometric space missions \corot\ and $Kepler$, such situations have been increasingly common. Most transiting candidates will have to be identified with additional means than the detection of the Keplerian motion, as already stated by, e.g. \citet{torres2011}. 

The full Bayesian analysis of the available data and models has been applied to \obj\ b using PASTIS \citep{diaz14}. This analysis has allowed quantifying the odds ratio between the planet scenarios and other false positive configurations, including hypothesis priors; it is larger than 1000, in strong support of the planet scenario. While some scenarios are rejected by the data, others are unfavored by hypothesis priors, and the consistent combination of these factors finally excludes false positive scenarios with a high level of confidence. More precisely, the role of 1) SED analyses (exclude PiB and TRIPLE),  2) adaptive optics imaging (combined with hypothesis priors, excludes BEB and BTP), and 3) radial velocity measurements (exclude undiluted binary, and constrain all other scenarios) has been major. The colored information of \corot\ offered little additional constraints for the resolution of the system and does not exclude one in particular. 

Interestingly, \obj\ b lies in the ''radius valley'' of the distribution of planetary radii (as known in May 2014). This valley corresponds to a gap in the 4-8 R$_\oplus$ range in the $Kepler$ sample \citep{howard12,marcy2014,rowe14}, although the detection is 98\% complete in the radius-period bin corresponding to \obj\ b \citep{petigura13}. Most of the objects that $Kepler$ has detected are smaller than 3 Earth radii. The period of \obj\ b, however, coincides with the peak period for transiting planets of less than 0.5 Jupiter radius; it is not an exotic object regarding its orbital distance, with respect with the population of planets known today.

In the family of planets confirmed or validated from the \corot\ mission, \obj\ b is the second smallest in size, after CoRoT-7 b. There are $\sim$ 20 additional candidates in this size range or smaller, that await further observations and analyses (Deleuil et al, in prep). For some of them, however, the magnitude of the star will prevent a scenario validation, because follow-up observations were not possible with current instrumentation, too time consuming, or not conclusive. In particular, adaptive optics imaging, that turned out to be critical in validating \obj\ b, is available, or gives strong enough constraints, for a subset of these candidates \citep{guenther13}. Also, even if radial-velocity observations were conducted on all promising candidates, and rejected grazing binaries in most cases (except fast rotators and hot stars), they usually give weak upper limits on the companion mass. Observed constraints differ from case to case, and only in a few cases are they significant enough to discard all false positives with strong confidence. PASTIS analyses will thus provide validation for only a handful of them (D\'{i}az et al, in prep). 

Table~\ref{starplanet_param_table} finally summarizes the parameters of the system. The radius of the candidate planet is 4.88$^{+0.17}_{-0.39}$ R$_\oplus$, while its mass is less than 49 M$_\oplus$ at 3-$\sigma$. This gives a density that is either compatible with the one of the Solar System gas giant planets, or, at maximum, between the density of Neptune and Mars. This parameter, and thus, the nature of the planet, is thus only loosely constrained. The planetary equilibrium temperature is $\sim$900K.

\begin{figure}
\centering
\includegraphics[width=0.5\textwidth,angle=0]{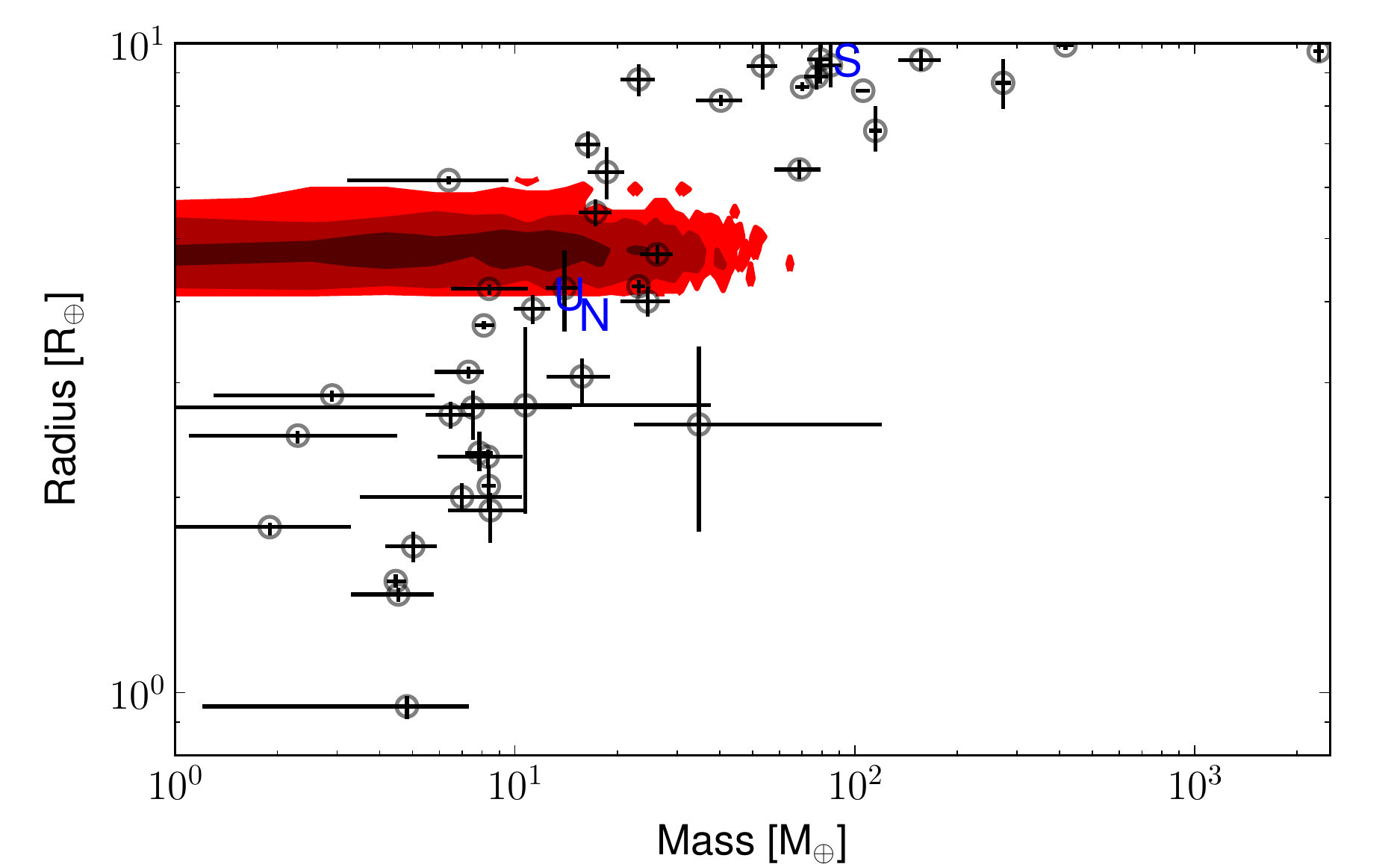}
\caption{Mass-radius of transiting extrasolar planets with measured radius and mass. Saturn, Uranus, and Neptune are labeled. The red shaded area shows \obj\ b.}
\label{fig13}
\end{figure}

Figure~\ref{fig13} shows how \obj\ b compares with other known transiting planets. In this radius range, other characterized planets tend to have a mass in the higher range of possible values for \obj\ b. These other planets in the radius range of \obj\ b either show TTVs (Kepler-11 and 18 systems) or have parent stars 4 magnitude brighter (HAT-P-11), which eases the RV observations. In order to derive the mass of \obj\ b in the future, intensive RV campaigns with Keck-HIRES and/or HARPS should be devoted, with a couple of measurements per night for one or two months. The future instrument VLT-ESPRESSO should also more easily detect the Keplerian motion of a few \ms\  due to such planets, thanks to a larger collecting area and exquisite RV precision. 
 
Modeling the internal structure of a giant planet without real mass constraint cannot be properly done. In addition, the radius of \obj\ b is close to the radius of Kepler-87 c and CoRoT-8 b, which have masses differing by a factor 10. Maybe the slightly enhanced metallicity of \obj\ ([Fe/H]=0.17) will direct the solution towards a large core mass, but only further RV observations will be able to settle to mean planet density, in a domain where very few planets are yet fully characterized. In the future,  TESS and PLATO missions will help discovering more planets in this radius/period range, and the ground-based complementary observations of the candidates will allow to get their mass with great accuracy, since the target stars will be much brighter than \corot\ typical targets. The underlying physics of this diversity in bulk densities will then be better understood.\\[0.5cm]

{\small \begin{table*}
\vspace{0cm}
\centering
\caption{Parameter prior distributions for all models.}            
\vspace{0cm}
\begin{minipage}[t]{20.0cm} 
\setlength{\tabcolsep}{3.0mm}
\renewcommand{\footnoterule}{}                          
\begin{tabular}{l c c }        
\hline\hline                 
\vspace{-0.25cm}\\
Parameter & Prior & Scenarios\\
\hline
Host/Target: & &  \\
Star density, $\rho_{\star}$ [$\rho_\odot$] & AN(0.49, 0.11, 0.27) & Planet\\
Star surface gravity, $\log$\,$g$ [cgs] &  N(4.3, 0.1)& BEB, BTP, PiB, Triple\\
Star effective temperature, $T_{\mathrm{eff}}$[K] &N(5780, 100) & All\\
Star metallicity, $[\rm{Fe/H}]$ [dex]  & N(0.21, 0.10)&All \\
Distance of the system, d [pc] &PL(2, 5, 6000) &All\\
Radial velocity linear drift [\ms year$^{-1}$] & U(-0.0001,0.0001)&All \\
$v$ sin$i_{\star}$ [\kms] & TUN(4.0,1.5, 0, 100)&All\\
Star albedo &U(0.6, 1.0) &All \\
\hline
Orbit: & &   \\
Planet orbital period, $P$ [days] & N(9.75607,0.00018)&All\\
Mid-transit/eclipse time, $T_{c}$ [BJD] &N(2454598.42020, 0.0014) &All \\
Orbital eccentricity, $e$  & U(0, 1)&All\\
Argument of periastron, $\omega$ [deg]  &U(0, 360) &All\\
Impact parameter, $b$ & U(0, 1) or U(0.0, 1.5)& Planet or BEB, BTP, PiB, Triple\\
\hline
Primary star: & & \\
Star mass, $M_\star$ [\Msun]                            & DPL(-1.55, -2.70, 1.0, 0.1, 30)& BEB, BTP, PiB, Triple \\
Age,  $t$ [$Gyr$]                              &U(6.3e-7, 13.8) &  BEB, BTP \\
Star metallicity, $[\rm{Fe/H}]$ [dex]   &  U(-2.5, 0.5)&  BEB, BTP\\
Distance of the system, d [pc] &  PL(2, 1, 20000) &  BEB, BTP \\
Star albedo &  U(0.6, 1.0)& BEB, BTP, PiB, Triple \\
v sini & U(0,30) &BEB, BTP, PiB, Triple \\
Systemic velocity & U(-200, 200)& BEB, BTP, PiB, Triple\\
\hline
Secondary star: & &  \\
Star mass, $M_\star$ [\Msun]              & DPL(-1.55,-2.70, 1.0, 0.1, 30) &BEB, Triple \\
Star albedo &U(0.6, 1.0) & BEB, Triple\\
$v$ sin$i_{\star}$ [\kms]   &U(0,30) &BEB, Triple\\
\hline
Planet: & &  \\
Albedo & U(0, 1)& Planet, BTP, PiB\\
Radial velocity semi-amplitude, $K$ [\ms]  & U(0.0, 0.1)& Planet \\
Radius ratio, $k=R_{p}/R_{\star}$  & J(1e-10, 0.1) & Planet\\
Planet mass, $M_{p}$ [M$_\oplus$ ]          & U(0, 90) & BTP, PiB \\
Planet radius, $R_{p}$[R$_\oplus$]             & J(0.09, 2.1)& BTP, PiB \\
\hline
Contamination CoRoT-W  [\%]   & N(0.019, 0.005)& All  \\
Normalization CoRoT-W   & U(0.999, 1.001)&All \\
Jitter CoRoT-W [mmag]  & U(0.0, 0.01)&All \\
Contamination CoRoT-R  [\%]    & N(0.010, 0.003)&All \\
Normalization CoRoT-R   & U(0.999, 1.001)&All \\
Jitter CoRoT-R [mmag] &U(0.0, 0.012) &All \\
Contamination CoRoT-G  [\%]   & N(0.041, 0.013)&All \\
Normalization CoRoT-G   & U(0.999, 1.001)&All\\
Jitter CoRoT-G [mmag] & U(0.0, 0.022)&All \\
Contamination CoRoT-B  [\%]  &N(0.062, 0.014) &All  \\
Normalization CoRoT-B  & U(0.999, 1.001)&All \\
Jitter CoRoT-B [mmag] &U(0.0, 0.023) &All\\
Offset HIRES, $\Delta RV_{HIRES}$ [\kms]  &U(-1.02, -0.9) & All\\
Jitter HIRES, $\sigma_{HIRES}$ [\ms]  & U(0.0, 0.034)& All\\
Offset HARPS, $\Delta RV_{HARPS}$ [\kms]   & U(-31.92, -31.86) &All\\
Jitter HARPS, $\sigma_{HARPS}$ [\ms]   & U(0.0, 0.068)& All \\
Offset SOPHIE, $\Delta RV_{SOPHIE}$ [\kms]   & U(-32.06, -31.75)&All \\
Jitter SOPHIE, $\sigma_{SOPHIE}$ [\ms]    & U(0, 0.1)& All\\
Jitter SED, $\sigma_{SED}$ [mag]  & U(0, 1)& All \\
\hline       
\vspace{-0.5cm}
\end{tabular}
\begin{list}{}{}
\tiny
\item U(xmin, xmax) = Uniform  distribution between xmin and xmax;
\item N(mu, sigma) = Normal distribution with mean $\mu$ and standard deviation $\sigma$
\item AN(mu, $\sigma -$, $\sigma +$) = Asymmetric normal distribution, with different widths at each side of mean value.
\item J(xmin, xmax) = Jeffreys distribution (log-flat) between xmin and xmax
\item TUN(mu, sigma, xmin, xmax) = Normal distribution in the interval [xmin, xmax]
\item PL(alpha, xmin, xmax) = power law with coefficient alpha, defined in the interval [xmin, xmax]
\item DPL(alpha, beta, x0, xmin, xmax) = A combination of two power law distributions, with coefficients alpha and beta, defined between [xmin, x0], and [x0, xmax], respectively. 
\item All distributions are normalised to unity.
\end{list}
\end{minipage}
\label{prior_table}  
\end{table*}
}


\small
\section*{Acknowledgments}
The authors are grateful to the referee, Tim Morton, who provided valuable input to our manuscript.
Some of the data presented are were obtained at the W.M. Keck Observatory from 
telescope time allocated to the National Aeronautics and Space Administration 
through the agency's scientific partnership with the California Institute of 
Technology and the University of California. The Observatory was made pos- 
sible by the generous financial support of the W.M. Keck Foundation.
This work is partially based on observations made with the IAC80 
telescope on the island of Tenerife in the Spanish 
Observatorio del Teide. 
The authors acknowledge DLR 
grants 50OW0204, 50OW0603, and 50QP0701; the French National Research Agency (ANR-08- JCJC-0102-01);
the ESA PRODEX program and the Swiss National Science Foundation; CNRS/CNES grant 07/0879-Corot; STFC grant 
ST/G002266; NASA Origins of Solar Systems grant  NNX09AB30G; and grant 
AYA2010-20982-C02-02 of the Spanish Ministry of Science and Innovation (MICINN); the European Research Council/European Community under the FP7 through Starting Grant agreement number 239953.

\bibliographystyle{mn2e}

\begin{thebibliography}{}

\bibitem[\protect\citeauthoryear{{Allard}, {Homeier} \& {Freytag}}{{Allard}
  et~al.}{2012}]{2012RSPTA.370.2765A}
{Allard} F.,  {Homeier} D.,    {Freytag} B.,  2012, Royal Society of London
  Philosophical Transactions Series A, 370, 2765

\bibitem[\protect\citeauthoryear{{Am{\^o}res} \& {L{\'e}pine}}{{Am{\^o}res} \&
  {L{\'e}pine}}{2005}]{2005AJ....130..659A}
{Am{\^o}res} E.~B.,  {L{\'e}pine} J.~R.~D.,  2005, AJ, 130, 659

\bibitem[\protect\citeauthoryear{{Auvergne}, {Bodin}, {Boisnard}, {Buey},
  {Chaintreuil}, {Epstein}, {Jouret}, {Lam-Trong}, {Levacher}, {Magnan},
  {Perez}, {Plasson}, {Plesseria}, {Peter}, {Steller}, {Tiph{\`e}ne}, {Baglin},
  {Agogu{\'e}}, {Appourchaux}, {Barbet]{auvergne09}
{Auvergne} M.,  {Bodin} P.,  {Boisnard} L.,  {Buey} J.-T.,  {Chaintreuil} S.,
  {Epstein} G.,  {Jouret} M.,  {Lam-Trong} T.,  {Levacher} P.,  {Magnan} A.,
  {Perez} R.,  {Plasson} P.,  {Plesseria} J.,  {Peter} G.,  {Steller} M.,
  {Tiph{\`e}ne} D.,  {Baglin} A.,  {Agogu{\'e}} P.,  {Appourchaux} T.,
  {Barbet} D.,  {Beaufort} T.,  {Bellenger} R.,  {Berlin} R.,  {Bernardi} P.,
  {Blouin} D.,  {Boumier} P.,  {Bonneau} F.,  {Briet} R.,  {Butler} B.,
  {Cautain} R.,  {Chiavassa} F.,  {Costes} V.,  {Cuvilho} J.,  {Cunha-Parro}
  V.,  {de Oliveira Fialho} F.,  {Decaudin} M.,  {Defise} J.-M.,  {Djalal} S.,
  {Docclo} A.,  {Drummond} R.,  {Dupuis} O.,  {Exil} G.,  {Faur{\'e}} C.,
  {Gaboriaud} A.,  {Gamet} P.,  {Gavalda} P.,  {Grolleau} E.,  {Gueguen} L.,
  {Guivarc'h} V.,  {Guterman} P.,  {Hasiba} J.,  {Huntzinger} G.,  {Hustaix}
  H.,  {Imbert} C.,  {Jeanville} G.,  {Johlander} B.,  {Jorda} L.,  {Journoud}
  P.,  {Karioty} F.,  {Kerjean} L.,  {Lafond} L.,  {Lapeyrere} V.,  {Landiech}
  P.,  {Larqu{\'e}} T.,  {Laudet} P.,  {Le Merrer} J.,  {Leporati} L.,
  {Leruyet} B.,  {Levieuge} B.,  {Llebaria} A.,  {Martin} L.,  {Mazy} E.,
  {Mesnager} J.-M.,  {Michel} J.-P.,  {Moalic} J.-P.,  {Monjoin} W.,  {Naudet}
  D.,  {Neukirchner} S.,  {Nguyen-Kim} K.,  {Ollivier} M.,  {Orcesi} J.-L.,
  {Ottacher} H.,  {Oulali} A.,  {Parisot} J.,  {Perruchot} S.,  {Piacentino}
  A.,  {Pinheiro da Silva} L.,  {Platzer} J.,  {Pontet} B.,  {Pradines} A.,
  {Quentin} C.,  {Rohbeck} U.,  {Rolland} G.,  {Rollenhagen} F.,  {Romagnan}
  R.,  {Russ} N.,  {Samadi} R.,  {Schmidt} R.,  {Schwartz} N.,  {Sebbag} I.,
  {Smit} H.,  {Sunter} W.,  {Tello} M.,  {Toulouse} P.,  {Ulmer} B.,
  {Vandermarcq} O.,  {Vergnault} E.,  {Wallner} R.,  {Waultier} G.,
  {Zanatta} P.,  2009, A\&A, 506, 411

\bibitem[\protect\citeauthoryear{{Batalha}, {Borucki}, {Bryson}, {Buchhave},
  {Caldwell}, {Christensen-Dalsgaard}, {Ciardi}, {Dunham}, {Fressin}, {Gautier}
  III, {Gilliland}, {Haas}, {Howell}, {Jenkins}, {Kjeldsen}, {Koch}, {Latham},
  {Lissauer}, {Marcy}, {Rowe}]{batalha10}
{Batalha} N.~M.,  {Borucki} W.~J.,  {Bryson} S.~T.,  {Buchhave} L.~A.,
  {Caldwell} D.~A.,  {Christensen-Dalsgaard} J.,  {Ciardi} D.,  {Dunham} E.~W.,
   {Fressin} F.,  {Gautier} III T.~N.,  {Gilliland} R.~L.,  {Haas} M.~R.,
  {Howell} S.~B.,  {Jenkins} J.~M.,  {Kjeldsen} H.,  {Koch} D.~G.,  {Latham}
  D.~W.,  {Lissauer} J.~J.,  {Marcy} G.~W.,  {Rowe} J.~F.,  {Sasselov} D.~D.,
  {Seager} S.,  {Steffen} J.~H.,  {Torres} G.,  {Basri} G.~S.,  {Brown} T.~M.,
  {Charbonneau} D.,  {Christiansen} J.,  {Clarke} B.,  {Cochran} W.~D.,
  {Dupree} A.,  {Fabrycky} D.~C.,  {Fischer} D.,  {Ford} E.~B.,  {Fortney} J.,
  {Girouard} F.~R.,  {Holman} M.~J.,  {Johnson} J.,  {Isaacson} H.,  {Klaus}
  T.~C.,  {Machalek} P.,  {Moorehead} A.~V.,  {Morehead} R.~C.,  {Ragozzine}
  D.,  {Tenenbaum} P.,  {Twicken} J.,  {Quinn} S.,  {VanCleve} J.,  {Walkowicz}
  L.~M.,  {Welsh} W.~F.,  {Devore} E.,    {Gould} A.,  2011, ApJ, 729, 27

\bibitem[\protect\citeauthoryear{{Bonfils}, {Delfosse}, {Udry}, {Forveille},
  {Mayor}, {Perrier}, {Bouchy}, {Gillon}, {Lovis}, {Pepe}, {Queloz}, {Santos},
  {S{\'e}gransan} \& {Bertaux}}{{Bonfils} et~al.}{2013}]{bonfils2013}
{Bonfils} X.,  {Delfosse} X.,  {Udry} S.,  {Forveille} T.,  {Mayor} M.,
  {Perrier} C.,  {Bouchy} F.,  {Gillon} M.,  {Lovis} C.,  {Pepe} F.,  {Queloz}
  D.,  {Santos} N.~C.,  {S{\'e}gransan} D.,    {Bertaux} J.-L.,  2013, A\&A,
  549, A109

\bibitem[\protect\citeauthoryear{{Bord{\'e}}, {Bouchy}, {Deleuil}, {Cabrera},
  {Jorda}, {Lovis}, {Csizmadia}, {Aigrain}, {Almenara}, {Alonso}, {Auvergne},
  {Baglin}, {Barge}, {Benz}, {Bonomo}, {Bruntt}, {Carone}, {Carpano}, {Deeg},
  {Dvorak}, {Erikson}, {Ferraz-]{borde2010}
{Bord{\'e}} P.,  {Bouchy} F.,  {Deleuil} M.,  {Cabrera} J.,  {Jorda} L.,
  {Lovis} C.,  {Csizmadia} S.,  {Aigrain} S.,  {Almenara} J.~M.,  {Alonso} R.,
  {Auvergne} M.,  {Baglin} A.,  {Barge} P.,  {Benz} W.,  {Bonomo} A.~S.,
  {Bruntt} H.,  {Carone} L.,  {Carpano} S.,  {Deeg} H.,  {Dvorak} R.,
  {Erikson} A.,  {Ferraz-Mello} S.,  {Fridlund} M.,  {Gandolfi} D.,  {Gazzano}
  J.-C.,  {Gillon} M.,  {Guenther} E.,  {Guillot} T.,  {Guterman} P.,  {Hatzes}
  A.,  {Havel} M.,  {H{\'e}brard} G.,  {Lammer} H.,  {L{\'e}ger} A.,  {Mayor}
  M.,  {Mazeh} T.,  {Moutou} C.,  {P{\"a}tzold} M.,  {Pepe} F.,  {Ollivier} M.,
   {Queloz} D.,  {Rauer} H.,  {Rouan} D.,  {Samuel} B.,  {Santerne} A.,
  {Schneider} J.,  {Tingley} B.,  {Udry} S.,  {Weingrill} J.,    {Wuchterl} G.,
   2010, A\&A, 520, A66+

\bibitem[\protect\citeauthoryear{{Borucki}, {Koch}, {Basri}, {Batalha}, {Boss},
  {Brown}, {Caldwell}, {Christensen-Dalsgaard}, {Cochran}, {DeVore}, {Dunham},
  {Dupree}, {Gautier} III, {Geary}, {Gilliland}, {Gould}, {Howell}, {Jenkins},
  {Kjeldsen}, {Latham}, {Li]{borucki}
{Borucki} W.~J.,  {Koch} D.~G.,  {Basri} G.,  {Batalha} N.,  {Boss} A.,
  {Brown} T.~M.,  {Caldwell} D.,  {Christensen-Dalsgaard} J.,  {Cochran} W.~D.,
   {DeVore} E.,  {Dunham} E.~W.,  {Dupree} A.~K.,  {Gautier} III T.~N.,
  {Geary} J.~C.,  {Gilliland} R.,  {Gould} A.,  {Howell} S.~B.,  {Jenkins}
  J.~M.,  {Kjeldsen} H.,  {Latham} D.~W.,  {Lissauer} J.~J.,  {Marcy} G.~W.,
  {Monet} D.~G.,  {Sasselov} D.,  {Tarter} J.,  {Charbonneau} D.,  {Doyle} L.,
  {Ford} E.~B.,  {Fortney} J.,  {Holman} M.~J.,  {Seager} S.,  {Steffen} J.~H.,
   {Welsh} W.~F.,  {Allen} C.,  {Bryson} S.~T.,  {Buchhave} L.,
  {Chandrasekaran} H.,  {Christiansen} J.~L.,  {Ciardi} D.,  {Clarke} B.~D.,
  {Dotson} J.~L.,  {Endl} M.,  {Fischer} D.,  {Fressin} F.,  {Haas} M.,
  {Horch} E.,  {Howard} A.,  {Isaacson} H.,  {Kolodziejczak} J.,  {Li} J.,
  {MacQueen} P.,  {Meibom} S.,  {Prsa} A.,  {Quintana} E.~V.,  {Rowe} J.,
  {Sherry} W.,  {Tenenbaum} P.,  {Torres} G.,  {Twicken} J.~D.,  {Van Cleve}
  J.,  {Walkowicz} L.,    {Wu} H.,  2011, ApJ, 728, 117

\bibitem[\protect\citeauthoryear{{Bruntt}, {Deleuil}, {Fridlund}, {Alonso},
  {Bouchy}, {Hatzes}, {Mayor}, {Moutou} \& {Queloz}}{{Bruntt}
  et~al.}{2010}]{bruntt}
{Bruntt} H.,  {Deleuil} M.,  {Fridlund} M.,  {Alonso} R.,  {Bouchy} F.,
  {Hatzes} A.,  {Mayor} M.,  {Moutou} C.,    {Queloz} D.,  2010, A\&A, 519,
  A51+

\bibitem[\protect\citeauthoryear{{Claret}}{{Claret}}{2001}]{2001MNRAS.327..989C}
{Claret} A.,  2001, MNRAS, 327, 989

\bibitem[\protect\citeauthoryear{{Claret} \& {Bloemen}}{{Claret} \&
  {Bloemen}}{2011}]{2011A&A...529A..75C}
{Claret} A.,  {Bloemen} S.,  2011, A\&A, 529, A75

\bibitem[\protect\citeauthoryear{{Deeg}, {Gillon}, {Shporer}, {Rouan},
  {Stecklum}, {Aigrain}, {Alapini}, {Almenara}, {Alonso}, {Barbieri}, {Bouchy},
  {Eisl{\"o}ffel}, {Erikson}, {Fridlund}, {Eigm{\"u}ller}, {Handler}, {Hatzes},
  {Kabath}, {Lendl}, {Mazeh}, {Mou]{deeg2009}
{Deeg} H.~J.,  {Gillon} M.,  {Shporer} A.,  {Rouan} D.,  {Stecklum} B.,
  {Aigrain} S.,  {Alapini} A.,  {Almenara} J.~M.,  {Alonso} R.,  {Barbieri} M.,
   {Bouchy} F.,  {Eisl{\"o}ffel} J.,  {Erikson} A.,  {Fridlund} M.,
  {Eigm{\"u}ller} P.,  {Handler} G.,  {Hatzes} A.,  {Kabath} P.,  {Lendl} M.,
  {Mazeh} T.,  {Moutou} C.,  {Queloz} D.,  {Rauer} H.,  {Rabus} M.,  {Tingley}
  B.,    {Titz} R.,  2009, A\&A, 506, 343

\bibitem[\protect\citeauthoryear{{Demory}, {Gillon}, {Deming}, {Valencia},
  {Seager}, {Benneke}, {Lovis}, {Cubillos}, {Harrington}, {Stevenson}, {Mayor},
  {Pepe}, {Queloz}, {S{\'e}gransan} \& {Udry}}{{Demory}
  et~al.}{2011}]{demory11}
{Demory} B.-O.,  {Gillon} M.,  {Deming} D.,  {Valencia} D.,  {Seager} S.,
  {Benneke} B.,  {Lovis} C.,  {Cubillos} P.,  {Harrington} J.,  {Stevenson}
  K.~B.,  {Mayor} M.,  {Pepe} F.,  {Queloz} D.,  {S{\'e}gransan} D.,    {Udry}
  S.,  2011, A\&A, 533, A114

\bibitem[\protect\citeauthoryear{{D{\'{\i}}az}, {Almenara}, {Santerne},
  {Moutou}, {Lethuillier} \& {Deleuil}}{{D{\'{\i}}az} et~al.}{2014}]{diaz14}
{D{\'{\i}}az} R.~F.,  {Almenara} J.~M.,  {Santerne} A.,  {Moutou} C.,
  {Lethuillier} A.,    {Deleuil} M.,  2014, MNRAS, 441, 983

\bibitem[\protect\citeauthoryear{{D{\'{\i}}az}, {Damiani}, {Deleuil},
  {Almenara}, {Moutou}, {Barros}, {Bonomo}, {Bouchy}, {Bruno}, {H{\'e}brard},
  {Montagnier} \& {Santerne}}{{D{\'{\i}}az} et~al.}{2013}]{2013A&A...551L...9D}
{D{\'{\i}}az} R.~F.,  {Damiani} C.,  {Deleuil} M.,  {Almenara} J.~M.,  {Moutou}
  C.,  {Barros} S.~C.~C.,  {Bonomo} A.~S.,  {Bouchy} F.,  {Bruno} G.,
  {H{\'e}brard} G.,  {Montagnier} G.,    {Santerne} A.,  2013, A\&A, 551, L9

\bibitem[\protect\citeauthoryear{{Dotter}, {Chaboyer}, {Jevremovi{\'c}},
  {Kostov}, {Baron} \& {Ferguson}}{{Dotter} et~al.}{2008}]{2008ApJS..178...89D}
{Dotter} A.,  {Chaboyer} B.,  {Jevremovi{\'c}} D.,  {Kostov} V.,  {Baron} E.,
   {Ferguson} J.~W.,  2008, ApJs, 178, 89

\bibitem[\protect\citeauthoryear{{Endl}, {K{\"u}rster} \& {Els}}{{Endl}
  et~al.}{2000}]{endl2000}
{Endl} M.,  {K{\"u}rster} M.,    {Els} S.,  2000, A\&A, 362, 585

\bibitem[\protect\citeauthoryear{{Etzel}}{{Etzel}}{1981}]{1981psbs.conf..111E}
{Etzel} P.~B.,  1981, in {Carling} E.~B.,  {Kopal} Z.,  eds, Photometric and
  Spectroscopic Binary Systems {A Simple Synthesis Method for Solving the
  Elements of Well-Detached Eclipsing Systems}.
p.~111

\bibitem[\protect\citeauthoryear{{Fitzpatrick}}{{Fitzpatrick}}{1999}]{1999PASP..111...63F}
{Fitzpatrick} E.~L.,  1999, PASP, 111, 63

\bibitem[\protect\citeauthoryear{{Fressin}, {Torres}, {Charbonneau}, {Bryson},
  {Christiansen}, {Dressing}, {Jenkins}, {Walkowicz} \& {Batalha}}{{Fressin}
  et~al.}{2013}]{fressin2013}
{Fressin} F.,  {Torres} G.,  {Charbonneau} D.,  {Bryson} S.~T.,  {Christiansen}
  J.,  {Dressing} C.~D.,  {Jenkins} J.~M.,  {Walkowicz} L.~M.,    {Batalha}
  N.~M.,  2013, ApJ, 766, 81

\bibitem[\protect\citeauthoryear{{Fressin}, {Torres}, {D{\'e}sert},
  {Charbonneau}, {Batalha}, {Fortney}, {Rowe}, {Allen}, {Borucki}, {Brown},
  {Bryson}, {Ciardi}, {Cochran}, {Deming}, {Dunham}, {Fabrycky}, {Gautier} III,
  {Gilliland}, {Henze}, {Holman}, {Howell]{fressin2011}
{Fressin} F.,  {Torres} G.,  {D{\'e}sert} J.-M.,  {Charbonneau} D.,  {Batalha}
  N.~M.,  {Fortney} J.~J.,  {Rowe} J.~F.,  {Allen} C.,  {Borucki} W.~J.,
  {Brown} T.~M.,  {Bryson} S.~T.,  {Ciardi} D.~R.,  {Cochran} W.~D.,  {Deming}
  D.,  {Dunham} E.~W.,  {Fabrycky} D.~C.,  {Gautier} III T.~N.,  {Gilliland}
  R.~L.,  {Henze} C.~E.,  {Holman} M.~J.,  {Howell} S.~B.,  {Jenkins} J.~M.,
  {Kinemuchi} K.,  {Knutson} H.,  {Koch} D.~G.,  {Latham} D.~W.,  {Lissauer}
  J.~J.,  {Marcy} G.~W.,  {Ragozzine} D.,  {Sasselov} D.~D.,  {Still} M.,
  {Tenenbaum} P.,    {Uddin} K.,  2011, ApJS, 197, 5

\bibitem[\protect\citeauthoryear{{Guenther}, {Fridlund}, {Alonso}, {Carpano},
  {Deeg}, {Deleuil}, {Dreizler}, {Endl}, {Gandolfi}, {Gillon}, {Guillot},
  {Jehin}, {L{\'e}ger}, {Moutou}, {Nortmann}, {Rouan}, {Samuel}, {Schneider} \&
  {Tingley}}{{Guenther} et~al.}{2]{guenther13}
{Guenther} E.~W.,  {Fridlund} M.,  {Alonso} R.,  {Carpano} S.,  {Deeg} H.~J.,
  {Deleuil} M.,  {Dreizler} S.,  {Endl} M.,  {Gandolfi} D.,  {Gillon} M.,
  {Guillot} T.,  {Jehin} E.,  {L{\'e}ger} A.,  {Moutou} C.,  {Nortmann} L.,
  {Rouan} D.,  {Samuel} B.,  {Schneider} J.,    {Tingley} B.,  2013, A\&A, 556,
  A75

\bibitem[\protect\citeauthoryear{{Holman}, {Fabrycky}, {Ragozzine}, {Ford},
  {Steffen}, {Welsh}, {Lissauer}, {Latham}, {Marcy}, {Walkowicz}, {Batalha},
  {Jenkins}, {Rowe}, {Cochran}, {Fressin}, {Torres}, {Buchhave}, {Sasselov},
  {Borucki}, {Koch}, {Basri}, {Brow]{holman}
{Holman} M.~J.,  {Fabrycky} D.~C.,  {Ragozzine} D.,  {Ford} E.~B.,  {Steffen}
  J.~H.,  {Welsh} W.~F.,  {Lissauer} J.~J.,  {Latham} D.~W.,  {Marcy} G.~W.,
  {Walkowicz} L.~M.,  {Batalha} N.~M.,  {Jenkins} J.~M.,  {Rowe} J.~F.,
  {Cochran} W.~D.,  {Fressin} F.,  {Torres} G.,  {Buchhave} L.~A.,  {Sasselov}
  D.~D.,  {Borucki} W.~J.,  {Koch} D.~G.,  {Basri} G.,  {Brown} T.~M.,
  {Caldwell} D.~A.,  {Charbonneau} D.,  {Dunham} E.~W.,  {Gautier} T.~N.,
  {Geary} J.~C.,  {Gilliland} R.~L.,  {Haas} M.~R.,  {Howell} S.~B.,  {Ciardi}
  D.~R.,  {Endl} M.,  {Fischer} D.,  {F{\"u}r{\'e}sz} G.,  {Hartman} J.~D.,
  {Isaacson} H.,  {Johnson} J.~A.,  {MacQueen} P.~J.,  {Moorhead} A.~V.,
  {Morehead} R.~C.,    {Orosz} J.~A.,  2010, Science, 330, 51

\bibitem[\protect\citeauthoryear{{Howard}, {Marcy}, {Bryson}, {Jenkins},
  {Rowe}, {Batalha}, {Borucki}, {Koch}, {Dunham}, {Gautier} III, {Van Cleve},
  {Cochran}, {Latham}, {Lissauer}, {Torres}, {Brown}, {Gilliland}, {Buchhave},
  {Caldwell}, {Christensen-Dalsgaar]{howard12}
{Howard} A.~W.,  {Marcy} G.~W.,  {Bryson} S.~T.,  {Jenkins} J.~M.,  {Rowe}
  J.~F.,  {Batalha} N.~M.,  {Borucki} W.~J.,  {Koch} D.~G.,  {Dunham} E.~W.,
  {Gautier} III T.~N.,  {Van Cleve} J.,  {Cochran} W.~D.,  {Latham} D.~W.,
  {Lissauer} J.~J.,  {Torres} G.,  {Brown} T.~M.,  {Gilliland} R.~L.,
  {Buchhave} L.~A.,  {Caldwell} D.~A.,  {Christensen-Dalsgaard} J.,  {Ciardi}
  D.,  {Fressin} F.,  {Haas} M.~R.,  {Howell} S.~B.,  {Kjeldsen} H.,  {Seager}
  S.,  {Rogers} L.,  {Sasselov} D.~D.,  {Steffen} J.~H.,  {Basri} G.~S.,
  {Charbonneau} D.,  {Christiansen} J.,  {Clarke} B.,  {Dupree} A.,  {Fabrycky}
  D.~C.,  {Fischer} D.~A.,  {Ford} E.~B.,  {Fortney} J.~J.,  {Tarter} J.,
  {Girouard} F.~R.,  {Holman} M.~J.,  {Johnson} J.~A.,  {Klaus} T.~C.,
  {Machalek} P.,  {Moorhead} A.~V.,  {Morehead} R.~C.,  {Ragozzine} D.,
  {Tenenbaum} P.,  {Twicken} J.~D.,  {Quinn} S.~N.,  {Isaacson} H.,  {Shporer}
  A.,  {Lucas} P.~W.,  {Walkowicz} L.~M.,  {Welsh} W.~F.,  {Boss} A.,  {Devore}
  E.,  {Gould} A.,  {Smith} J.~C.,  {Morris} R.~L.,  {Prsa} A.,  {Morton}
  T.~D.,  {Still} M.,  {Thompson} S.~E.,  {Mullally} F.,  {Endl} M.,
  {MacQueen} P.~J.,  2012, ApJs, 201, 15

\bibitem[\protect\citeauthoryear{Kass \& Raftery}{Kass \&
  Raftery}{1995}]{Kass:1995:BF}
Kass R.~E.,  Raftery A.~E.,  1995, Journal of the American Statistical
  Association, 90, 773

\bibitem[\protect\citeauthoryear{{L{\'e}ger}, {Rouan}, {Schneider}, {Barge},
  {Fridlund}, {Samuel}, {Ollivier}, {Guenther}, {Deleuil}, {Deeg}, {Auvergne},
  {Alonso}, {Aigrain}, {Alapini}, {Almenara}, {Baglin}, {Barbieri}, {Bruntt},
  {Bord{\'e}}, {Bouchy}, {Cabre]{leger09}
{L{\'e}ger} A.,  {Rouan} D.,  {Schneider} J.,  {Barge} P.,  {Fridlund} M.,
  {Samuel} B.,  {Ollivier} M.,  {Guenther} E.,  {Deleuil} M.,  {Deeg} H.~J.,
  {Auvergne} M.,  {Alonso} R.,  {Aigrain} S.,  {Alapini} A.,  {Almenara} J.~M.,
   {Baglin} A.,  {Barbieri} M.,  {Bruntt} H.,  {Bord{\'e}} P.,  {Bouchy} F.,
  {Cabrera} J.,  {Catala} C.,  {Carone} L.,  {Carpano} S.,  {Csizmadia} S.,
  {Dvorak} R.,  {Erikson} A.,  {Ferraz-Mello} S.,  {Foing} B.,  {Fressin} F.,
  {Gandolfi} D.,  {Gillon} M.,  {Gondoin} P.,  {Grasset} O.,  {Guillot} T.,
  {Hatzes} A.,  {H{\'e}brard} G.,  {Jorda} L.,  {Lammer} H.,  {Llebaria} A.,
  {Loeillet} B.,  {Mayor} M.,  {Mazeh} T.,  {Moutou} C.,  {P{\"a}tzold} M.,
  {Pont} F.,  {Queloz} D.,  {Rauer} H.,  {Renner} S.,  {Samadi} R.,  {Shporer}
  A.,  {Sotin} C.,  {Tingley} B.,  {Wuchterl} G.,  {Adda} M.,  {Agogu} P.,
  {Appourchaux} T.,  {Ballans} H.,  {Baron} P.,  {Beaufort} T.,  {Bellenger}
  R.,  {Berlin} R.,  {Bernardi} P.,  {Blouin} D.,  {Baudin} F.,  {Bodin} P.,
  {Boisnard} L.,  {Boit} L.,  {Bonneau} F.,  {Borzeix} S.,  {Briet} R.,  {Buey}
  J.-T.,  {Butler} B.,  {Cailleau} D.,  {Cautain} R.,  {Chabaud} P.-Y.,
  {Chaintreuil} S.,  {Chiavassa} F.,  {Costes} V.,  {Cuna Parrho} V.,  {de
  Oliveira Fialho} F.,  {Decaudin} M.,  {Defise} J.-M.,  {Djalal} S.,
  {Epstein} G.,  {Exil} G.-E.,  {Faur{\'e}} C.,  {Fenouillet} T.,  {Gaboriaud}
  A.,  {Gallic} A.,  {Gamet} P.,  {Gavalda} P.,  {Grolleau} E.,  {Gruneisen}
  R.,  {Gueguen} L.,  {Guis} V.,  {Guivarc'h} V.,  {Guterman} P.,  {Hallouard}
  D.,  {Hasiba} J.,  {Heuripeau} F.,  {Huntzinger} G.,  {Hustaix} H.,  {Imad}
  C.,  {Imbert} C.,  {Johlander} B.,  {Jouret} M.,  {Journoud} P.,  {Karioty}
  F.,  {Kerjean} L.,  {Lafaille} V.,  {Lafond} L.,  {Lam-Trong} T.,  {Landiech}
  P.,  {Lapeyrere} V.,  {Larqu{\'e}} T.,  {Laudet} P.,  {Lautier} N.,  {Lecann}
  H.,  {Lefevre} L.,  {Leruyet} B.,  {Levacher} P.,  {Magnan} A.,  {Mazy} E.,
  {Mertens} F.,  {Mesnager} J.-M.,  {Meunier} J.-C.,  {Michel} J.-P.,
  {Monjoin} W.,  {Naudet} D.,  {Nguyen-Kim} K.,  {Orcesi} J.-L.,  {Ottacher}
  H.,  {Perez} R.,  {Peter} G.,  {Plasson} P.,  {Plesseria} J.-Y.,  {Pontet}
  B.,  {Pradines} A.,  {Quentin} C.,  {Reynaud} J.-L.,  {Rolland} G.,
  {Rollenhagen} F.,  {Romagnan} R.,  {Russ} N.,  {Schmidt} R.,  {Schwartz} N.,
  {Sebbag} I.,  {Sedes} G.,  {Smit} H.,  {Steller} M.~B.,  {Sunter} W.,
  {Surace} C.,  {Tello} M.,  {Tiph{\`e}ne} D.,  {Toulouse} P.,  {Ulmer} B.,
  {Vandermarcq} O.,  {Vergnault} E.,  {Vuillemin} A.,    {Zanatta} P.,  2009,
  A\&A, 506, 287

\bibitem[\protect\citeauthoryear{{Lissauer}, {Fabrycky}, {Ford}, {Borucki},
  {Fressin}, {Marcy}, {Orosz}, {Rowe}, {Torres}, {Welsh}, {Batalha}, {Bryson},
  {Buchhave}, {Caldwell}, {Carter}, {Charbonneau}, {Christiansen}, {Cochran},
  {Desert}, {Dunham}, {Fanelli},]{lissauer}
{Lissauer} J.~J.,  {Fabrycky} D.~C.,  {Ford} E.~B.,  {Borucki} W.~J.,
  {Fressin} F.,  {Marcy} G.~W.,  {Orosz} J.~A.,  {Rowe} J.~F.,  {Torres} G.,
  {Welsh} W.~F.,  {Batalha} N.~M.,  {Bryson} S.~T.,  {Buchhave} L.~A.,
  {Caldwell} D.~A.,  {Carter} J.~A.,  {Charbonneau} D.,  {Christiansen} J.~L.,
  {Cochran} W.~D.,  {Desert} J.-M.,  {Dunham} E.~W.,  {Fanelli} M.~N.,
  {Fortney} J.~J.,  {Gautier} III T.~N.,  {Geary} J.~C.,  {Gilliland} R.~L.,
  {Haas} M.~R.,  {Hall} J.~R.,  {Holman} M.~J.,  {Koch} D.~G.,  {Latham} D.~W.,
   {Lopez} E.,  {McCauliff} S.,  {Miller} N.,  {Morehead} R.~C.,  {Quintana}
  E.~V.,  {Ragozzine} D.,  {Sasselov} D.,  {Short} D.~R.,    {Steffen} J.~H.,
  2011, Nature, 470, 53

\bibitem[\protect\citeauthoryear{{Lissauer}, {Marcy}, {Bryson}, {Rowe},
  {Jontof-Hutter}, {Agol}, {Borucki}, {Carter}, {Ford}, {Gilliland}, {Kolbl},
  {Star}, {Steffen} \& {Torres}}{{Lissauer} et~al.}{2014}]{lissauer14}
{Lissauer} J.~J.,  {Marcy} G.~W.,  {Bryson} S.~T.,  {Rowe} J.~F.,
  {Jontof-Hutter} D.,  {Agol} E.,  {Borucki} W.~J.,  {Carter} J.~A.,  {Ford}
  E.~B.,  {Gilliland} R.~L.,  {Kolbl} R.,  {Star} K.~M.,  {Steffen} J.~H.,
  {Torres} G.,  2014, ApJ, 784, 44

\bibitem[\protect\citeauthoryear{{Marcy}, {Isaacson}, {Howard}, {Rowe},
  {Jenkins}, {Bryson}, {Latham}, {Howell}, {Gautier} III, {Batalha}, {Rogers},
  {Ciardi}, {Fischer}, {Gilliland}, {Kjeldsen}, {Christensen-Dalsgaard},
  {Huber}, {Chaplin}, {Basu}, {Buchhave},]{marcy2014}
{Marcy} G.~W.,  {Isaacson} H.,  {Howard} A.~W.,  {Rowe} J.~F.,  {Jenkins}
  J.~M.,  {Bryson} S.~T.,  {Latham} D.~W.,  {Howell} S.~B.,  {Gautier} III
  T.~N.,  {Batalha} N.~M.,  {Rogers} L.,  {Ciardi} D.,  {Fischer} D.~A.,
  {Gilliland} R.~L.,  {Kjeldsen} H.,  {Christensen-Dalsgaard} J.,  {Huber} D.,
  {Chaplin} W.~J.,  {Basu} S.,  {Buchhave} L.~A.,  {Quinn} S.~N.,  {Borucki}
  W.~J.,  {Koch} D.~G.,  {Hunter} R.,  {Caldwell} D.~A.,  {Van Cleve} J.,
  {Kolbl} R.,  {Weiss} L.~M.,  {Petigura} E.,  {Seager} S.,  {Morton} T.,
  {Johnson} J.~A.,  {Ballard} S.,  {Burke} C.,  {Cochran} W.~D.,  {Endl} M.,
  {MacQueen} P.,  {Everett} M.~E.,  {Lissauer} J.~J.,  {Ford} E.~B.,  {Torres}
  G.,  {Fressin} F.,  {Brown} T.~M.,  {Steffen} J.~H.,  {Charbonneau} D.,
  {Basri} G.~S.,  {Sasselov} D.~D.,  {Winn} J.,  {Sanchis-Ojeda} R.,
  {Christiansen} J.,  {Adams} E.,  {Henze} C.,  {Dupree} A.,  {Fabrycky} D.~C.,
   {Fortney} J.~J.,  {Tarter} J.,  {Holman} M.~J.,  {Tenenbaum} P.,  {Shporer}
  A.,  {Lucas} P.~W.,  {Welsh} W.~F.,  {Orosz} J.~A.,  {Bedding} T.~R.,
  {Campante} T.~L.,  {Davies} G.~R.,  {Elsworth} Y.,  {Handberg} R.,  {Hekker}
  S.,  {Karoff} C.,  {Kawaler} S.~D.,  {Lund} M.~N.,  {Lundkvist} M.,
  {Metcalfe} T.~S.,  {Miglio} A.,  {Silva Aguirre} V.,  {Stello} D.,  {White}
  T.~R.,  {Boss} A.,  {Devore} E.,  {Gould} A.,  {Prsa} A.,  {Agol} E.,
  {Barclay} T.,  {Coughlin} J.,  {Brugamyer} E.,  {Mullally} F.,  {Quintana}
  E.~V.,  {Still} M.,  {Thompson} S.~E.,  {Morrison} D.,  {Twicken} J.~D.,
  {D{\'e}sert} J.-M.,  {Carter} J.,  {Crepp} J.~R.,  {H{\'e}brard} G.,
  {Santerne} A.,  {Moutou} C.,  {Sobeck} C.,  {Hudgins} D.,  {Haas} M.~R.,
  {Robertson} P.,  {Lillo-Box} J.,    {Barrado} D.,  2014, ApJs, 210, 20

\bibitem[\protect\citeauthoryear{{Mayor}, {Pepe}, {Queloz}, {Bouchy},
  {Rupprecht}, {Lo Curto}, {Avila}, {Benz}, {Bertaux}, {Bonfils}, {Dall},
  {Dekker}, {Delabre}, {Eckert}, {Fleury}, {Gilliotte}, {Gojak}, {Guzman},
  {Kohler}, {Lizon}, {Longinotti}, {Lovis}, {M]{mayor2003}
{Mayor} M.,  {Pepe} F.,  {Queloz} D.,  {Bouchy} F.,  {Rupprecht} G.,  {Lo
  Curto} G.,  {Avila} G.,  {Benz} W.,  {Bertaux} J.-L.,  {Bonfils} X.,  {Dall}
  T.,  {Dekker} H.,  {Delabre} B.,  {Eckert} W.,  {Fleury} M.,  {Gilliotte} A.,
   {Gojak} D.,  {Guzman} J.~C.,  {Kohler} D.,  {Lizon} J.-L.,  {Longinotti} A.,
   {Lovis} C.,  {Megevand} D.,  {Pasquini} L.,  {Reyes} J.,  {Sivan} J.-P.,
  {Sosnowska} D.,  {Soto} R.,  {Udry} S.,  {van Kesteren} A.,  {Weber} L.,
  {Weilenmann} U.,  2003, The Messenger, 114, 20

\bibitem[\protect\citeauthoryear{{Morton}}{{Morton}}{2012}]{morton12}
{Morton} T.~D.,  2012, ApJ, 761, 6

\bibitem[\protect\citeauthoryear{{Moutou}, {Deleuil}, {Guillot}, {Baglin},
  {Bord{\'e}}, {Bouchy}, {Cabrera}, {Csizmadia} \& {Deeg}}{{Moutou}
  et~al.}{2013}]{moutou13}
{Moutou} C.,  {Deleuil} M.,  {Guillot} T.,  {Baglin} A.,  {Bord{\'e}} P.,
  {Bouchy} F.,  {Cabrera} J.,  {Csizmadia} S.,    {Deeg} H.~J.,  2013, Icarus,
  226, 1625

\bibitem[\protect\citeauthoryear{{Nelson} \& {Davis}}{{Nelson} \&
  {Davis}}{1972}]{1972ApJ...174..617N}
{Nelson} B.,  {Davis} W.~D.,  1972, ApJ, 174, 617

\bibitem[\protect\citeauthoryear{{Petigura}, {Marcy} \& {Howard}}{{Petigura}
  et~al.}{2013}]{petigura13}
{Petigura} E.~A.,  {Marcy} G.~W.,    {Howard} A.~W.,  2013, ApJ, 770, 69

\bibitem[\protect\citeauthoryear{{Popper} \& {Etzel}}{{Popper} \&
  {Etzel}}{1981}]{1981AJ.....86..102P}
{Popper} D.~M.,  {Etzel} P.~B.,  1981, AJ, 86, 102

\bibitem[\protect\citeauthoryear{{Raghavan}, {McAlister}, {Henry}, {Latham},
  {Marcy}, {Mason}, {Gies}, {White} \& {ten Brummelaar}}{{Raghavan}
  et~al.}{2010}]{raghavan2010}
{Raghavan} D.,  {McAlister} H.~A.,  {Henry} T.~J.,  {Latham} D.~W.,  {Marcy}
  G.~W.,  {Mason} B.~D.,  {Gies} D.~R.,  {White} R.~J.,    {ten Brummelaar}
  T.~A.,  2010, ApJs, 190, 1

\bibitem[\protect\citeauthoryear{{Robin}, {Reyl{\'e}}, {Derri{\`e}re} \&
  {Picaud}}{{Robin} et~al.}{2003}]{robin2003}
{Robin} A.~C.,  {Reyl{\'e}} C.,  {Derri{\`e}re} S.,    {Picaud} S.,  2003,
  A\&A, 409, 523

\bibitem[\protect\citeauthoryear{{Rowe}, {Bryson}, {Marcy}, {Lissauer},
  {Jontof-Hutter}, {Mullally}, {Gilliland}, {Issacson}, {Ford}, {Howell},
  {Borucki}, {Haas}, {Huber}, {Steffen}, {Thompson}, {Quintana}, {Barclay},
  {Still}, {Fortney}, {Gautier} III, {Hunte]{rowe14}
{Rowe} J.~F.,  {Bryson} S.~T.,  {Marcy} G.~W.,  {Lissauer} J.~J.,
  {Jontof-Hutter} D.,  {Mullally} F.,  {Gilliland} R.~L.,  {Issacson} H.,
  {Ford} E.,  {Howell} S.~B.,  {Borucki} W.~J.,  {Haas} M.,  {Huber} D.,
  {Steffen} J.~H.,  {Thompson} S.~E.,  {Quintana} E.,  {Barclay} T.,  {Still}
  M.,  {Fortney} J.,  {Gautier} III T.~N.,  {Hunter} R.,  {Caldwell} D.~A.,
  {Ciardi} D.~R.,  {Devore} E.,  {Cochran} W.,  {Jenkins} J.,  {Agol} E.,
  {Carter} J.~A.,    {Geary} J.,  2014, ApJ, 784, 45

\bibitem[\protect\citeauthoryear{{Skrutskie}, {Cutri}, {Stiening}, {Weinberg},
  {Schneider}, {Carpenter}, {Beichman}, {Capps}, {Chester}, {Elias}, {Huchra},
  {Liebert}, {Lonsdale}, {Monet}, {Price}, {Seitzer}, {Jarrett}, {Kirkpatrick},
  {Gizis}, {Howard}, {Evans]{2006AJ....131.1163S}
{Skrutskie} M.~F.,  {Cutri} R.~M.,  {Stiening} R.,  {Weinberg} M.~D.,
  {Schneider} S.,  {Carpenter} J.~M.,  {Beichman} C.,  {Capps} R.,  {Chester}
  T.,  {Elias} J.,  {Huchra} J.,  {Liebert} J.,  {Lonsdale} C.,  {Monet} D.~G.,
   {Price} S.,  {Seitzer} P.,  {Jarrett} T.,  {Kirkpatrick} J.~D.,  {Gizis}
  J.~E.,  {Howard} E.,  {Evans} T.,  {Fowler} J.,  {Fullmer} L.,  {Hurt} R.,
  {Light} R.,  {Kopan} E.~L.,  {Marsh} K.~A.,  {McCallon} H.~L.,  {Tam} R.,
  {Van Dyk} S.,    {Wheelock} S.,  2006, AJ, 131, 1163

\bibitem[\protect\citeauthoryear{{Southworth}}{{Southworth}}{2011}]{2011MNRAS.417.2166S}
{Southworth} J.,  2011, MNRAS, 417, 2166

\bibitem[\protect\citeauthoryear{{Torres}, {Fressin}, {Batalha}, {Borucki},
  {Brown}, {Bryson}, {Buchhave}, {Charbonneau}, {Ciardi}, {Dunham}, {Fabrycky},
  {Ford}, {Gautier} III, {Gilliland}, {Holman}, {Howell}, {Isaacson},
  {Jenkins}, {Koch}, {Latham}, {Lissaue]{torres2011}
{Torres} G.,  {Fressin} F.,  {Batalha} N.~M.,  {Borucki} W.~J.,  {Brown} T.~M.,
   {Bryson} S.~T.,  {Buchhave} L.~A.,  {Charbonneau} D.,  {Ciardi} D.~R.,
  {Dunham} E.~W.,  {Fabrycky} D.~C.,  {Ford} E.~B.,  {Gautier} III T.~N.,
  {Gilliland} R.~L.,  {Holman} M.~J.,  {Howell} S.~B.,  {Isaacson} H.,
  {Jenkins} J.~M.,  {Koch} D.~G.,  {Latham} D.~W.,  {Lissauer} J.~J.,  {Marcy}
  G.~W.,  {Monet} D.~G.,  {Prsa} A.,  {Quinn} S.~N.,  {Ragozzine} D.,  {Rowe}
  J.~F.,  {Sasselov} D.~D.,  {Steffen} J.~H.,    {Welsh} W.~F.,  2011, ApJ,
  727, 24

\bibitem[\protect\citeauthoryear{{Torres}, {Konacki}, {Sasselov} \&
  {Jha}}{{Torres} et~al.}{2005}]{torres2005}
{Torres} G.,  {Konacki} M.,  {Sasselov} D.~D.,    {Jha} S.,  2005, ApJ, 619,
  558

\bibitem[\protect\citeauthoryear{{Tuomi} \& {Jones}}{{Tuomi} \&
  {Jones}}{2012}]{2012A&A...544A.116T}
{Tuomi} M.,  {Jones} H.~R.~A.,  2012, A\&A, 544, A116

\bibitem[\protect\citeauthoryear{{Winn}, {Matthews}, {Dawson}, {Fabrycky},
  {Holman}, {Kallinger}, {Kuschnig}, {Sasselov}, {Dragomir}, {Guenther},
  {Moffat}, {Rowe}, {Rucinski} \& {Weiss}}{{Winn} et~al.}{2011}]{winn11}
{Winn} J.~N.,  {Matthews} J.~M.,  {Dawson} R.~I.,  {Fabrycky} D.,  {Holman}
  M.~J.,  {Kallinger} T.,  {Kuschnig} R.,  {Sasselov} D.,  {Dragomir} D.,
  {Guenther} D.~B.,  {Moffat} A.~F.~J.,  {Rowe} J.~F.,  {Rucinski} S.,
  {Weiss} W.~W.,  2011, ApJL, 737, L18

\bibitem[\protect\citeauthoryear{{Wright}, {Eisenhardt}, {Mainzer}, {Ressler},
  {Cutri}, {Jarrett}, {Kirkpatrick}, {Padgett}, {McMillan}, {Skrutskie},
  {Stanford}, {Cohen}, {Walker}, {Mather}, {Leisawitz}, {Gautier} III,
  {McLean}, {Benford}, {Lonsdale}, {Blain}]{2010AJ....140.1868W}
{Wright} E.~L.,  {Eisenhardt} P.~R.~M.,  {Mainzer} A.~K.,  {Ressler} M.~E.,
  {Cutri} R.~M.,  {Jarrett} T.,  {Kirkpatrick} J.~D.,  {Padgett} D.,
  {McMillan} R.~S.,  {Skrutskie} M.,  {Stanford} S.~A.,  {Cohen} M.,  {Walker}
  R.~G.,  {Mather} J.~C.,  {Leisawitz} D.,  {Gautier} III T.~N.,  {McLean} I.,
  {Benford} D.,  {Lonsdale} C.~J.,  {Blain} A.,  {Mendez} B.,  {Irace} W.~R.,
  {Duval} V.,  {Liu} F.,  {Royer} D.,  {Heinrichsen} I.,  {Howard} J.,
  {Shannon} M.,  {Kendall} M.,  {Walsh} A.~L.,  {Larsen} M.,  {Cardon} J.~G.,
  {Schick} S.,  {Schwalm} M.,  {Abid} M.,  {Fabinsky} B.,  {Naes} L.,    {Tsai}
  C.-W.,  2010, AJ, 140, 1868

\end{thebibliography}


\begin{thebibliography}{99}


\bibitem[\protect\citeauthoryear{Allard, Homeier and Freytag} {2012}]{allard} Allard, F., Homeier, D. and Freytag, B., 2012, Royal Society of London Philosophical Transactions Series A, 370, 2765
\bibitem[\protect\citeauthoryear{Am{\^o}res and L{\'e}pine} {2005}]{amores} Am{\^o}res, E.~B. and L{\'e}pine, J.~R.~D., 2005, AJ 130, 659
\bibitem[\protect\citeauthoryear{Auvergne et al} {2009}]{auvergne09} Auvergne, M., Bodin, P., Boisnard, L. et al, 2009, A\&A 506, 411
\bibitem[\protect\citeauthoryear{Batalha et al} {2011}]{batalha10} Batalha, N., Borucki, W.J., Bryson, S.T. et al, 2011, ApJ 729, 27
\bibitem[\protect\citeauthoryear{Bonfils et al} {2013}]{bonfils2013} Bonfils, X., Delfosse, X., Udry, S. et al, 2013, A\&A 549, A109
\bibitem[\protect\citeauthoryear{Borucki et al} {2011}]{borucki} Borucki, W.J., Koch, D.G., Basri, G et al, 2011, ApJ 728, 117
\bibitem[\protect\citeauthoryear{Bord\'e et al} {2010}]{borde2010} Bord\'e, P., Bouchy, F., Deleuil, M. et al, 2010, A\&A 520, A66
\bibitem[\protect\citeauthoryear{Borucki et al} {2011}]{borucki} Borucki, W.J., Koch, D.G., Basri, G et al, 2011, ApJ 728, 117
\bibitem[\protect\citeauthoryear{Bruntt et al} {2010}]{bruntt} Bruntt, H., Deleuil, M., Fridlund, M. et al, 2010, A\&A 519, A51
\bibitem[\protect\citeauthoryear{Claret} {2001}]{claret01} Claret, A., 2001, A\&A 327, 989
\bibitem[\protect\citeauthoryear{Claret \& Bloemen} {2011}]{claret} Claret, A., and Bloemen, S., 2011, A\&A 529, A75
\bibitem[\protect\citeauthoryear{Deeg et al} {2009}]{deeg2009} Deeg, H.J., Gillon, M., Shporer, A. et al, 2009, A\&A 506, 343
\bibitem[\protect\citeauthoryear{Demory et al} {2011}]{demory11} Demory, B.-O., Gillon, M., Deming, D. et al, 2011, A\&A 533, A114
\bibitem[\protect\citeauthoryear{D\'{i}az et al} {2014}]{diaz14} D\'{i}az, R.F., Almenara, J.M., Santerne, A. et al, 2014, MNRAS 441, 983
\bibitem[\protect\citeauthoryear{D\'{i}az et al} {2013}]{diaz13} D\'{i}az, R.F., Daminai, C., Deleuil, M. et al, 2013, A\&A 551, L9
\bibitem[\protect\citeauthoryear{Dotter et al} {2008}]{dotter} Dotter, A., Chaboyer, B., Jevremovi{\'c}, D. et al, 2008, ApJs 178, 89
\bibitem[\protect\citeauthoryear{Endl et al} {2000}]{endl2000} Endl, M., K\"urster, M. and Els, S., 2000, A\&A 362, 585
\bibitem[\protect\citeauthoryear{Etzel} {1981}]{etzel} Etzel, P.B., 1981, in {\it Photometric and Spectroscopic Binary Systems}, eds Carling \& Kopal
\bibitem[\protect\citeauthoryear{Fitzpatrick} {1999}]{fitzpatrick} Fitzpatrick, E.~L., 1999, PASP 111, 63
\bibitem[\protect\citeauthoryear{Fressin et al} {2011}]{fressin2011} Fressin, F., Torres, G., D\'esert, J.-M. et al, 2011, ApJS 197, 5
\bibitem[\protect\citeauthoryear{Fressin et al} {2013}]{fressin2013} Fressin, F., Torres, G., Charbonneau, D. et al, 2013, ApJ 766, 81
\bibitem[\protect\citeauthoryear{Guenther et al} {2013}]{guenther13} Guenther, E.W., Fridlund, M., Alonso, R. et al, 2013, A\&A 556, A75
\bibitem[\protect\citeauthoryear{Holman et al} {2010}]{holman} Holman, M.J., Fabrycky, D.C., Ragozzine, D. et al, 2010, Science 330, 51
\bibitem[\protect\citeauthoryear{Howard et al} {2012}]{howard12} Howard, A.W., Marcy, G.W., Bryson, S.T. et al, 2012, ApJS 201, 15
\bibitem[\protect\citeauthoryear{Kass and Raftery} {1995}]{Kass} Kass, R.E. and Raftery, A.E., 1995, JSTNAL 90, 773
\bibitem[\protect\citeauthoryear{L\'eger et al} {2009}]{leger09} L\'eger, A., Rouan, D., Schneider, J. et al, 2009, A\&A  506, 287
\bibitem[\protect\citeauthoryear{Lissauer et al} {2011}]{lissauer} Lissauer, J.J., Fabrycky, D.C., Ford, E.B et al, 2011,Nature, 470, 53
\bibitem[\protect\citeauthoryear{Lissauer et al} {2014}]{lissauer14} Lissauer, J.~J., Marcy, G.~W. Bryson, S.~T. et al, 2014, ApJ 784, 44
\bibitem[\protect\citeauthoryear{Marcy et al} {2014}]{marcy2014} Marcy, G.W., Isaacson, H., Howard, A.W. et al, 2014, ApJs 210, 20
\bibitem[\protect\citeauthoryear{Mayor et al} {2003}]{mayor2003} Mayor, M., Pepe, F., Queloz, D. et al, 2003, A\&A 114, 20
\bibitem[\protect\citeauthoryear{Morton} {2012}]{morton12} Morton, T. D., 2012, ApJ 761, 6
\bibitem[\protect\citeauthoryear{Moutou et al} {2013}]{moutou13} Moutou, C., Deleuil, M., Guillot, T. et al, 2013, Icarus 226, 1625
\bibitem[\protect\citeauthoryear{Nelson \& Davies} {1972}]{nelson} Nelson, B. and Davies, W.D., 1972, ApJ 174, 617
\bibitem[\protect\citeauthoryear{Petigura et al} {2013}]{petigura13} Petigura, E.~A. and Marcy, G.~W. and Howard, A.~W., 2013, ApJ
\bibitem[\protect\citeauthoryear{Popper \& Etzel} {1981}]{popper} Popper, D.M. and Etzel, P.B., 1981, AJ 86, 102
\bibitem[\protect\citeauthoryear{Raghavan et al} {2010}]{raghavan2010} Raghavan, D., McAlister, H.A., Henry, T.J. et al 2010, ApJS 190, 1
\bibitem[\protect\citeauthoryear{Rappaport et al} {2013}]{rappaport13} Rappaport, S., Deck, K., Levine, A. et al, 2013, ApJ 768, 33
\bibitem[\protect\citeauthoryear{Robin et al} {2003}]{robin2003} Robin, A.C., Reyl\'e, C., Derri\`ere, S. \& Picaud, S., 2003, A\&A 409, 523
\bibitem[\protect\citeauthoryear{Rowe et al} {2014}]{rowe14} Rowe, J.F., Bryson, S.T., Marcy, G.W. et al, 2014, ApJ 784, 45
\bibitem[\protect\citeauthoryear{Skrutskie et al} {2006}]{skrutskie} Skrutskie, M.F., Cutri, R.M., Stiening, R. et al 2006, AJ 131, 1163
\bibitem[\protect\citeauthoryear{Southworth} {2011}]{southworth} Southworth, J., 2011, MNRAS 417, 2166
\bibitem[\protect\citeauthoryear{Torres et al} {2005}]{torres2005} Torres, G., Konacki, M., Sasselov, D.D. and Jha, S., 2005, ApJ 619, 558
\bibitem[\protect\citeauthoryear{Torres et al} {2011}]{torres2011} Torres, G., Fressin, F., Batalha, N.M. et al, 2011, ApJ 727, 24
\bibitem[\protect\citeauthoryear{Tuomi \& Jones} {2012}]{tuomi} Tuomi, M. and Jones, H.~R.~A., 2012, A\&A 544, 116
\bibitem[\protect\citeauthoryear{Winn et al} {2011}]{winn11} Winn, J.N., Matthews, J.M., Dawson, R.I. et al, 2011, ApJ 737, 18
\bibitem[\protect\citeauthoryear{Wright et al} {2010}]{wright} Wright, E.L., Eisenhardt, P.R.M., Mainzer, A.K. et al, 2010, AJ 140, 1868



\end{thebibliography}

\label{lastpage}
\end{document}